\documentclass{IEEEtran}

\ifCLASSINFOpdf
  \usepackage[pdftex]{graphicx}
  \DeclareGraphicsExtensions{.pdf,.jpeg,.png}
\else
  \usepackage[dvips]{graphicx}
\fi
\usepackage{bm}
\usepackage{threeparttable}
\usepackage{epstopdf}
\usepackage[compress,nospace]{cite}
\usepackage{amsthm}
\usepackage{stfloats}
\usepackage[cmex10]{amsmath}
\usepackage{amsmath}
\usepackage{cases}
\usepackage{siunitx}
\usepackage{amssymb}
\usepackage{wasysym}
\usepackage[ruled,linesnumbered]{algorithm2e}
\usepackage[colorlinks,linkcolor=blue,anchorcolor=blue,citecolor=blue,bookmarks=blue]{hyperref}
\usepackage{array}
\usepackage{url}
\usepackage{color}
\usepackage{subfigure}
\usepackage{verbatim}
\usepackage{multicol}  
\usepackage{multirow} 
\usepackage{booktabs}
\usepackage{makecell}
\usepackage{upgreek}
\allowdisplaybreaks[4]
\usepackage{setspace}
\begin{document}

%% Title and Authors
\title{Entropy-based Probing Beam Selection and Beam Prediction via Deep Learning}

\author{Fan~Meng, Cheng~Zhang,~\IEEEmembership{Member,~IEEE}, \\Yongming~Huang,~\IEEEmembership{Senior Member,~IEEE}, Zhilei~Zhang, Xiaoyu~Bai, Zhaohua~Lu
	\thanks{This work was supported in part by the National Key R\&D Program of China under Grant 2020YFB1806600 and the National Natural Science Foundation of China under Grant No. 62225107, 62001103 and 62201394, and the Fundamental Research Funds for the Central Universities under Grant 2242022k60002. Part of this work has been submitted for possible presentation at the IEEE International Communications Conference (ICC): Wireless Communications Symposium, Denver, United States, June 2024 \cite{icc24}. (Corresponding authors: Y.\ Huang and C.\ Zhang)}
	\thanks{F. Meng, Z. Zhang, X. Bai, C. Zhang and Y. Huang are with the Purple Mountain Laboratories, Nanjing 211111, China (e-mail: mengfan@pmlabs.com.cn; zhangzhilei@pmlabs.com.cn; baixiaoyu@pmlabs.com.cn; zhangcheng\_seu@seu.edu.cn; huangym@seu.edu.cn). C. Zhang and Y. Huang are also with the National Mobile Communications Research Laboratory, School of Information Science and Engineering, Southeast University, Nanjing 210096, China. Z. Lu is with the ZTE Corporation and State Key Laboratory of Mobile Network and Mobile Multimedia Technology, Shenzhen 518057, China (e-mail: lu.zhaohua@zte.com.cn).}
}

% The paper headers
\markboth{Submitted to IEEE Transactions on Communications}
{Shell \MakeLowercase{\textit{et al.}}: Bare Demo of IEEEtran.cls for Journals}

\maketitle

\begin{abstract}

Hierarchical beam search in mmWave communications incurs substantial training overhead, necessitating deep learning-enabled beam predictions to effectively leverage channel priors and mitigate this overhead. In this study, we introduce a comprehensive probabilistic model of power distribution in beamspace, and formulate the joint optimization problem of probing beam selection and probabilistic beam prediction as an entropy minimization problem. Then, we propose a greedy scheme to iteratively and alternately solve this problem, where a transformer-based beam predictor is trained to estimate the conditional power distribution based on the probing beams and user location within each iteration, and the trained predictor selects an unmeasured beam that minimizes the entropy of remaining beams. To further reduce the number of interactions and the computational complexity of the iterative scheme, we propose a two-stage probing beam selection scheme. Firstly, probing beams are selected from a location-specific codebook designed by an entropy-based criterion, and predictions are made with corresponding feedback. Secondly, the optimal beam is identified using additional probing beams with the highest predicted power values. Simulation results demonstrate the superiority of the proposed schemes compared to hierarchical beam search and beam prediction with uniform probing beams.

\end{abstract}

\begin{IEEEkeywords}
	
mmWave communication, beam prediction, probing beam selection, deep learning, entropy minimization.

\end{IEEEkeywords}

\section{Introduction}\label{sec:introduction}

In B5G/6G wireless communications, millimeter wave (mmWave) communication is emerging as an appealing solution to provide abundant available spectrum to meet the critical demands of exploding data traffic~\cite{7959169}. However, the high path loss of mmWave signals poses a significant challenge to data transmission, resulting in limited coverage area. The small carrier wavelength enables packing a large number of antenna elements into small form factors. Leveraging the large antenna arrays employed at the transmitter and receiver, mmWave systems perform directional beamforming~\cite{6600706} to overcome the high free-space path loss of mmWave signals and also reduce spatial interference. Nevertheless, the massive antennas bring significant challenges for channel estimation, beam alignment/tracking (BA/T), especially in highly mobile and/or complex environments such as high-speed railway, unmanned aerial vehicle (UAV), urban macro (UMa).

Pencil-like beamforming for mmWave scenarios such as UMa is challenging. Environmental factors, i.e., wind flow and precipitation, moving vehicles and pedestrians, can cause drastic variations in received power fluctuation. As the transceiver units of mmWave base stations (BS) are mounted to facilities such as poles, pillars or street lamps, vibration and movement can cause unacceptable outage probability if BA/T is not frequently performed~\cite{6600706}. Meanwhile, link stability is improved with frequent BA/T, but at the cost of high beam training overheads. Therefore, low-overhead mmWave BA/T while maintaining link stability is essential.

Traditional model-driven BA/T including exhaustive and hierarchical beam search~\cite{5262295, 5379063, 7390101}, fail to adequately exploit the (partial) channel state information (CSI) prior, resulting in large overheads. Furthermore, the CSI prior is difficult to analytically characterize especially in complex scenarios, thus the model-driven methods inherently are inappropriate for the implicit CSI prior. Deep learning (DL)~\cite{Goodfellow2016Deep} has been identified as an enabling technology for future wireless mobile networks~\cite{8742579, 8715338}, and it has received extensive attention for precoding~\cite{9210124, 9562487, 10001369}, positioning~\cite{9390409, 9427239}, CSI compression and reconstruction~\cite{9513579, 9931713}, beam management~\cite{9069211, 9269463, 9598911, 9846951, 9954418}, network optimization\cite{9650869}. Data-driven or -aided BA/T is a promising technique that automatically learns and exploits the underlying correlations of CSI across different times, frequencies, spaces or other out-of-band information~\cite{8114345, 9923616}, to reduce CSI acquisition overhead and improve system spectral efficiency and robustness.

\subsection{Related Work}

Conventional BA/T is measurement-based, the transmit/receive beam is an element within the measured beam set, e.g., the beam with maximal received power is selected in a single-user link-level system. Predictive BA/T, on the other hand, derives the maximum received power or RSRP in the entire beamspace with few or no measurements. The main difference is that the selected beam is not necessarily measured in the beam prediction, but can be inferred from measured beams at different times, frequencies and spaces. In some special scenarios, even instantaneous measurement is not required and the beam prediction is performed only with out-of-band information such as location, motion and orientation of the mobile user (MU) ~\cite{9512417, 9598911, 9846951}.

Considering the time correlation, some studies establish the dynamics of the mmWave channel to realize beam prediction~\cite{9605254, 9944873, 9805463}. The studies in~\cite{9605254} propose a two-level probabilistic beam prediction. On a long-time scale, a variational auto-encoder uses noisy beam-training feedback to learn a probabilistic model of beam dynamics and enable predictive beam-tracking; on a short-time scale, an adaptive beam-training procedure is formulated as a partially observable Markov decision process and solved by reinforcement learning. In~\cite{9944873}, the neural ordinary differential equation is exploited to predict the arbitrary-instant optimal beam between the current and next beam training instants.

Due to the same physical environment, channels in low and high carrier frequencies are correlated. To reap this benefit, many studies learn to predict the mmWave channel with full sub-6 GHz channel and limited low-overhead measurement of the mmWave channel~\cite{9121328, 9529181, 9950519}. In~\cite{9529181}, a dual-input neural network (NN) architecture is designed to merge the sub-6 GHz channel and the mmWave channel of a few active antennas. In addition, an antenna selection method is introduced to better match the mmWave channel instead of a uniform pilot design. Moreover, the work in~\cite{9121328} considers blockage prediction which is important to proactively improve the link reliability.

Location-awareness is becoming a fundamental feature to support various mobile applications, and the radio access network is a promising infrastructure to efficiently and intelligently utilize wireless resources by integrating sensing and communication~\cite{8226757, 8246850, 9373011}. Therefore, recent studies estimate or directly use the MU location or geometric environment between transceivers as side information for beam prediction~\cite{8823977, 9369831, 9422201, 9439176, 9512417}. The Gaussian process (GP) is a probabilistic machine learning model that performs inference with uncertainties, and it has been well-applied in small-sample low-dimensional beam prediction~\cite{9422201, 9439176, 9512417, 9580358}. In mmWave fixed wireless access, researchers develop an explicit mapping between transmit/receive beams and MU physical coordinates via a GP~\cite{9422201}. Similarly in the mmWave UAV network, GP-enabled beam management scheme utilizing angular domain information is proposed~\cite{9439176, 9512417} to rapidly establish and reliably maintain the communication links. Small-sample learning is promising for real-time online adaption, but it inherently lacks the ability to exact complex priors from plenty of data. Moreover, GP training involves matrix inversion, making it difficult for large-sample high-dimensional problems where DL tools are better options~\cite{8823977, 9013296, 9129762, 9369831}. As a large amount of training data is presented, offline learning and online inference with DL is well-investigated. In~\cite{8823977}, a mapping from the user location to the beam pairs (fingerprints) is realized by a deep neural network (NN), with labeled data collected in different locations. Meanwhile, single spatial information is insufficient to accurately infer the reference signal receive power (RSRP) of narrow beams, and low-overhead probing is necessary~\cite{9690703}. 

\subsection{Motivation and Contribution}
 
Considering a mmWave system with massive antenna arrays, we investigate the location-aware probing beam selection and the probabilistic beam prediction problems. In general, we use the multivariate Gaussian model to approximate the distribution of RSRP in beamspace, and a beam predictor is to estimate the RSRP distribution with MU location. Given the learned beam predictor, the probing beam selection is to minimize the conditional entropy of the unmeasured beams, by selecting the training beam combination which is a subset of the discrete Fourier transform (DFT) codebook. The involved technical difficulties are as follows.
\begin{itemize}
	\item \textbf{Prediction Model}. The beamspace of RSRP with massive antennas is high-dimensional and the underlying channel prior is implicit and complex. Thus, model-driven or shallow data-driven schemes such as GP cannot work well. On the other hand, DL has the merit of efficiently extracting the channel prior, but the existing literature rarely considers probabilistic inference.
	\item \textbf{Uncertainty Evaluation}. Global uncertainty can be evaluated by entropy, but this is incompatible with beamforming which usually concerns only with the optimal beam with maximum RSRP. Meanwhile, the optimal beam can be in any direction, and local uncertainty cannot address this issue. Apparently, there is a trade-off between global and local uncertainty in the probing beam selection.
	\item \textbf{Computational Complexity}. The probing beam selection is a combinatorial optimization problem requiring exhaustive search, which has extremely high computational complexity in high-dimensional beamspace and is infeasible for practical use. Furthermore, each search consumes one beam prediction involving matrix inversion, which exacerbates this issue.
\end{itemize}

To address these difficulties, first to realize probabilistic inference, we propose a DL-based beam predictor which is trained by the maximum likelihood (ML) criterion, to estimate the conditional RSRP distribution with the MU location and probing beams. Second, to achieve a trade-off between global and local uncertainty, we mask the entropy with a weight matrix, which is designed by the predicted RSRP values. Third, to reduce the computational complexity, we design a greedy solution to iteratively solve the probing beam selection problem. Furthermore, to reduce the number of interactions and the computational complexity of the iterative solution, we propose a two-stage probing beam selection scheme. The two-level scheme is feasible for practical implementation, which has only two interactions and one beam prediction operation. Our technical contributions are summarized as follows.
\begin{itemize}
\item We establish a generic probabilistic model of the RSRP in beamspace, and formulate the joint probing beam selection and probabilistic beam prediction as an entropy minimization problem. To obtain a tradeoff between the overall entropy and the local entropy w.r.t. large-power beams, we extend the problem as a weighted entropy minimization with a tunable diagonal matrix.
\item We propose a greedy scheme, i.e., Iter-BP\&PBS, to iteratively solve the weighted entropy minimization problem, to reduce the computational complexity. During an iteration, a beam predictor is trained to estimate the conditional RSRP distribution with the probing beams and the MU location. The learned predictor then selects an unmeasured beam to minimize the weighted entropy of the remaining unmeasured beams.
\item We propose a two-stage probing beam selection scheme, i.e., 2S-BP\&PBS, to further reduce the number of interactions and the computational complexity of Iter-BP\&PBS. Firstly, the BS selects probing beams in a location-specific codebook by the MU location, and roughly locates the optimal beam by the prediction with the RSRP feedbacks. Secondly, the BS probes several beams with top-predicted RSRP values to find the optimal beam.
\item We design a scalable beam predictor composed of a mean network and a variance network, using the transformer as the backbone. We design an ML-based cost function to realize probabilistic inference, and simplify the variance network output as a diagonal covariance matrix, to further reduce computational complexity and achieve numerical stability.
\item Simulation results for an urban scenario demonstrate the superior performance of the proposed schemes compared to the existing hierarchical beam search and beam prediction with uniform probing beams. In addition, the two-stage scheme has low computational, storage and interaction requirements, which is important for real-time deployment.
\end{itemize}

The rest of this paper is organized as follows. The system model and problem formulation are described in Section~\ref{sec:system}. The iterative algorithm for probing beam selection and beam prediction is clarified in Section~\ref{sec:algorithm}, and the two-stage probing beam selection is introduced in Section~\ref{sec:two_stage}. Furthermore, the design of the transformer-based beam predictor is given in Section~\ref{sec:transformer}. The numerical results are shown in Section~\ref{sec:sim}, and the conclusions are drawn in Section~\ref{sec:conclusion}.

\textit{Notations}: We use lowercase (uppercase) boldface $ {\bf A}({\bf a}) $ to denote the matrix (vector), and $ a $ is a scalar. Calligraphy letter $ \mathcal{A} $ represents the set. Superscripts $ (\cdot)^{\mathsf{T}} $ represents the transpose. $ \mathrm{det} $, $ \textup{diag} $, $ |\cdot| $, $ \|\cdot\|_2 $, $ \otimes $ respectively denote determinant, diagonal, absolute, $ \ell_2 $ norm, and Kronecker product operators. $ \mathbb{E}\{\cdot\} $, $ \mathbb{R} $ and $ \mathbb{C} $ respectively represent the expectation, the real and complex fields.

\section{System Model and Problem Formulation}\label{sec:system}

\subsection{System Model}
	
We consider a link-level mmWave massive multiple-input single-output (MISO) communication system composed of one BS and one MU. The BS is equipped with a massive planer antenna array with $ N $ antennas connected to one radio frequency (RF) chain, and the MU is equipped with one isotropic antenna. The scenarios can be readily expanded to cellular or cell-free networks with multiple BSs and MUs, where the beam training is executed through time or frequency division.

\subsubsection{Channel Model}

Without loss of generality, the wireless mmWave propagation is characterized by multi-path propagation due to interactions (reflections, diffractions, penetrations, scattering) at stationary obstacles (hills, buildings, towers) and mobile objects (cars, pedestrians). According to the 3GPP channel modeling~\cite{38901}, the downlink mmWave channel $ {\bf h} \in \mathbb{C}^{N \times 1} $ is modeled as a combination of line-of-sight (LoS) and non-LoS (NLoS) channels, i.e.,
\begin{equation}\label{equ:H}
{\bf h} = {\bf h}_{\textup{LoS}} + {\bf h}_{\textup{NLoS}}.
\end{equation}
The LoS channel $ {\bf h}_{\textup{LoS}} $ has only a dominant path, the NLoS $ {\bf h}_{\textup{NLoS}} $ is consisting of $ N_{\textup{cl}} $ dominant clusters, and each cluster is composed of $ N_{\textup{ray}} $ rays. Thus, the narrow-band channel vectors in the antenna domain respectively are described as
\begin{align}\label{equ:H_element}
{\bf h}_{\textup{LoS}} & = \alpha_{\textup{LoS}} \boldsymbol{\psi}(\phi_{\textup{LoS}}, \theta_{\textup{LoS}}),\\
{\bf h}_{\textup{NLoS}} & = \sum_{m=1}^{N_{\textup{cl}}} \sum_{n=1}^{N_{\textup{ray}}} \alpha_{m,n} \boldsymbol{\psi}(\phi_{m,n}, \theta_{m,n}),
\end{align}
where $ \alpha $ is a complex channel gain, $ \phi $ and $ \theta $ respectively are the angles of departure in horizontal and vertical directions, $ \boldsymbol{\psi} $ is the planer array response at the BS which is given as follows
\begin{equation}\label{equ:array_response}
\boldsymbol{\psi}(\phi, \theta) = \boldsymbol{a}_{\textup{xy}}(\phi, \theta) \otimes \boldsymbol{a}_{\textup{z}}(\theta),
\end{equation}
where
\begin{align}\label{equ:array_response_xyz}
\boldsymbol{a}_{\textup{xy}}(\phi, \theta) & = \frac{1}{\sqrt{N_{\phi}}}[1, e^{\jmath\pi\sin\phi\sin\theta}, \cdots, e^{\jmath \pi (N_{\phi}-1) \sin\phi\sin\theta}]^{\mathsf{T}},\\
\boldsymbol{a}_{\textup{z}}(\theta) & = \frac{1}{\sqrt{N_{\theta}}}[1, e^{\jmath \pi\cos\theta}, \cdots, e^{\jmath \pi (N_{\theta}-1) \cos\theta}]^{\mathsf{T}},
\end{align}
where $ N_{\phi} $ and $ N_{\theta} $ respectively are the numbers of antennas in the horizontal and vertical dimensions, and $ N = N_{\phi} N_{\theta} $. The rays in cluster $ m $ are closely distributed around the center of this cluster in angles $ \{\phi_{m}, \theta_{m}\} $. Particularly, $ \alpha_{\textup{LoS}} = 0 $ indicates the LoS path is blocked.

\subsubsection{Data Model}

The term `data model' refers to the modeling and formulation of available data, specifically designed for training and evaluation purposes.

The BS exhaustively sweeps the beams in the DFT codebook $ {\bf A} \in \mathbb{C}^{N \times N} $, and the observed RSRP at the MU side is represented as 
\begin{equation}\label{equ:beam}
{\bf x} = |{\bf A h} + {\bf n}_{x}|^2,
\end{equation}
where $ {\bf n}_{x} \in \mathbb{C}^{N \times 1} $ is the measurement noise following $ \mathcal{N}({\bf 0}, \sigma_x^2 {\bf I}_N) $ where $ \sigma_x^2 $ is the noise variance. The relative position of the MU w.r.t. the BS is
\begin{equation}\label{equ:r_i} 
{\bf s} = {\bf s}_{\textup{r}} - {\bf s}_{\textup{t}} + {\bf n}_{s},
\end{equation}
where $ {\bf s}_{\textup{t}} \in \mathbb{R}^{2 \times 1} $ and $ {\bf s}_{\textup{r}} \in \mathbb{R}^{2 \times 1} $ respectively are 2D locations of the BS and the MU, $ {\bf n}_{s} $ denotes the positioning error following $ \mathcal{N}({\bf 0}, \sigma_s^2 {\bf I}_2) $.

Measurement-based BA/T typically consumes large beam training overhead to align the optimal beam. Using the potential mapping from the side information to the RSRP, can significantly reduce the overhead. In this work, we propose to realize BA/T with a small number of probing beams and the MU location, thus the design of probing beams is crucial. 
Concretely, given the relative MU location $ {\bf s} $ and the maximal number of probing beams $ L $, the BS sends $ L $ probing beams in $ {\bf A} $ and receives the counterpart MU feedbacks, then predicts the RSRPs of other beams in $ {\bf A} $.

\subsection{Problem Formulation}\label{sec:problem}

In this subsection, we formally articulate the problems of beam prediction and probing beam selection.

The RSRP $ {\bf x} $ follows an unknown multi-parameter distribution $ \Pr({\bf x}) $. For the convenience of analysis and implementation of learning methods, we assume that $ {\bf x} $ approximately follows the multivariate Gaussian distribution, i.e.,
\begin{equation}\label{equ:h_variable}
\mathcal{N}({\bf x}; \boldsymbol{\mu}, \boldsymbol{\Sigma}) = \frac{1}{\sqrt{\det (2 \pi\boldsymbol{\Sigma})}} \exp\Big(- \frac{1}{2} ({\bf x} - \boldsymbol{\mu}) \boldsymbol{\Sigma}^{-1} ({\bf x} - \boldsymbol{\mu})^{\mathsf{T}}\Big),
\end{equation}
where $ \boldsymbol{\mu} $ and $ \boldsymbol{\Sigma} $ respectively are the mean vector and covariance matrix of $ {\bf x} $, and they are regarded as a function w.r.t. the side information $ {\bf s} $.

Given $ {\bf s} $, the beam prediction problem is to estimate the distribution of $ {\bf x} $, which is modeled as a ML estimation, i.e.,
\begin{equation}\label{equ:mle}
\max_{\boldsymbol{\Theta}_f, \boldsymbol{\Theta}_g} \mathbb{E}_{{\bf x}, {\bf s}} \Big\{ \ln \mathcal{N}({\bf x}; \boldsymbol{\mu}, \boldsymbol{\Sigma}) \Big\},
\end{equation}
where
\begin{align}
\boldsymbol{\mu} & = f({\bf s}; \boldsymbol{\Theta}_f),\label{equ:f}\\
\boldsymbol{\Sigma} & = g({\bf s}; \boldsymbol{\Theta}_g),\label{equ:g}
\end{align}
where $ f $ and $ g $ respectively are the functions of the mean network and the variance network, $ \boldsymbol{\Theta}_f $ and $ \boldsymbol{\Theta}_g $ are the corresponding learnable parameters. Evidently, mean square error (MSE) minimization-based beam prediction w.r.t. $ {\bf x} $ is a special case of \eqref{equ:mle} with the simpler setting $ \boldsymbol{\Sigma} = {\bf I}_N $.

We define a set of measured beam indices $ \mathcal{Q} \subseteq \{1, \cdots, N\} $, and a set of un-measured beam indices $ \mathcal{P} \subseteq \{1, \cdots, N\} $. Given the number of measured beams $ |\mathcal{Q}| \leq L $, and we have $ \mathcal{P} \bigcup \mathcal{Q} = \{1, \cdots, N\} $ and $ \mathcal{P} \bigcap \mathcal{Q} = \varnothing $. Equivalently, $ \{x_i | i \in \mathcal{Q}\} $ and $ \{x_i | i \in \mathcal{P}\} $ respectively can be vectorized as $ {\bf x}_{\mathcal{Q}} $ and $ {\bf x}_{\mathcal{P}} $. We re-arrange the order of $ {\bf x} $ to get $ {\bf x} = [{\bf x}_{\mathcal{P}}^{\mathsf{T}}, {\bf x}_{\mathcal{Q}}^{\mathsf{T}}]^{\mathsf{T}} $, thus the counterpart mean and covariance respectively are $ \boldsymbol{\mu} = [\boldsymbol{\mu}_{\mathcal{P}}^{\mathsf{T}}, \boldsymbol{\mu}_{\mathcal{Q}}^{\mathsf{T}}]^{\mathsf{T}} $ and
\begin{align}
\boldsymbol{\Sigma} & = 
\begin{bmatrix}
\boldsymbol{\Sigma}_{\mathcal{P} \mathcal{P}} & \boldsymbol{\Sigma}_{\mathcal{P} \mathcal{Q}}\\
\boldsymbol{\Sigma}_{\mathcal{Q} \mathcal{P}} & \boldsymbol{\Sigma}_{\mathcal{Q} \mathcal{Q}}
\end{bmatrix},\nonumber
\end{align}
where $ {\bf x}_{\mathcal{P}} \sim \mathcal{N}(\boldsymbol{\mu}_{\mathcal{P}}, \boldsymbol{\Sigma}_{\mathcal{P} \mathcal{P}}) $ and $ {\bf x}_{\mathcal{Q}} \sim \mathcal{N}(\boldsymbol{\mu}_{\mathcal{Q}}, \boldsymbol{\Sigma}_{\mathcal{Q} \mathcal{Q}}) $. Given measured beams $ {\bf x}_{\mathcal{Q}} $, the distribution of the un-measured beams $ {\bf x}_{\mathcal{P}} $ is a conditional multivariate Gaussian variable~\cite{eaton1983multivariate} following $ \mathcal{N}({\bf x}_{\mathcal{P}}; \boldsymbol{\mu}_{\mathcal{P} | \mathcal{Q}}, \boldsymbol{\Sigma}_{\mathcal{P}|\mathcal{Q}}) $ where
\begin{align}
\boldsymbol{\mu}_{\mathcal{P} | \mathcal{Q}} & = \boldsymbol{\mu}_{\mathcal{P}} + \boldsymbol{\Sigma}_{\mathcal{P} \mathcal{Q}} \boldsymbol{\Sigma}_{\mathcal{Q} \mathcal{Q}}^{-1}({\bf{x}}_{\mathcal{Q}} - \boldsymbol{\mu}_{\mathcal{Q}}),\label{equ:close_1} \\
\boldsymbol{\Sigma}_{\mathcal{P}|\mathcal{Q}} & = \boldsymbol{\Sigma}_{\mathcal{P} \mathcal{P}} - \boldsymbol{\Sigma}_{\mathcal{P} \mathcal{Q}} \boldsymbol{\Sigma}_{\mathcal{Q} \mathcal{Q}}^{-1} \boldsymbol{\Sigma}_{\mathcal{Q} \mathcal{P}}.\label{equ:close_2}
\end{align}
Based on the learned statistics $ \boldsymbol{\mu}, \boldsymbol{\Sigma} $ in \eqref{equ:f} and \eqref{equ:g}, the conditional distribution $ \mathcal{N}({\bf x}_{\mathcal{P}}; \boldsymbol{\mu}_{\mathcal{P} | \mathcal{Q}}, \boldsymbol{\Sigma}_{\mathcal{P}|\mathcal{Q}}) $ has a closed-form expression based on  \eqref{equ:close_1} and \eqref{equ:close_2}.

The target of probing beam selection is to minimize the uncertainty of the unmeasured beams $ {\bf x}_{\mathcal{P}} $, by finding a combination of measured beams $ {\bf x}_{\mathcal{Q}} $. The conditional entropy of $ {\bf x}_{\mathcal{P}} $, i.e., 
\begin{equation}\label{equ:entropy}
\textup{H}({\bf x}_{\mathcal{P}}) = \frac{1}{2} \ln (2 \pi e \det \boldsymbol{\Sigma}_{\mathcal{P}|\mathcal{Q}}), 
\end{equation}
can perform as an uncertainty measure of $ {\bf x}_{\mathcal{P}} $. The entropy $ \textup{H}({\bf x}_{\mathcal{P}}) $ is monotonous w.r.t. $ \det \boldsymbol{\Sigma}_{\mathcal{P}|\mathcal{Q}} $. Given the maximal number of probing beams $ L $, the estimate statistics $ \boldsymbol{\mu} $ and $ \boldsymbol{\Sigma} $, the subsequent probing beam selection problem considers the minimization of conditional entropy $ \textup{H}({\bf x}_{\mathcal{P}}) $ w.r.t. the set of measured beam indices $ \mathcal{Q} $. This problem can be equivalently written as a combinatorial optimization problem as follows
\begin{equation}\label{equ:det_sigma}
\begin{split}
\min_{\mathcal{Q}} & \det \boldsymbol{\Sigma}_{\mathcal{P}|\mathcal{Q}}\\
\textup{s.t.} \,\, & |\mathcal{Q}| \leq L.
\end{split}
\end{equation}

\section{Iterative Beam Prediction and Probing Beam Selection}\label{sec:algorithm}

In Section~\ref{sec:problem} we have generally formulated the beam prediction and probing beam selection problems, and we will discuss the counterpart solutions below.

\subsection{Weighted Entropy Minimization}

In practice, most BA/T only considers the beams with high or maximum RSRP values, so we propose to rewrite problem \eqref{equ:det_sigma} as
\begin{equation}\label{equ:det_mask_sigma}
\begin{split}
\min_{\mathcal{Q}} & \det (\boldsymbol{\Delta}^{\frac{1}{2}}_{\mathcal{P}|\mathcal{Q}} \boldsymbol{\Sigma}_{\mathcal{P}|\mathcal{Q}} \boldsymbol{\Delta}^{\frac{1}{2}}_{\mathcal{P}|\mathcal{Q}})\\
\textup{s.t.} \,\, & |\mathcal{Q}| = L,
\end{split}
\end{equation}
where the diagonal matrix $ \boldsymbol{\Delta}_{\mathcal{P}|\mathcal{Q}} $ performs as a mask with its diagonal element indicating the weight. The minimum of \eqref{equ:det_sigma} is only obtained with $ |\mathcal{Q}| = L $, and \eqref{equ:det_sigma} is a special case of \eqref{equ:det_mask_sigma} with $ \boldsymbol{\Delta}_{\mathcal{P}|\mathcal{Q}} = {\bf I}_{N - |\mathcal{Q}|} $. We propose a mask function $ h $ w.r.t. the corresponding conditional mean, i.e., 
\begin{equation}\label{equ:mask_re}
\boldsymbol{\Delta}_{\mathcal{P}|\mathcal{Q}} = h(\boldsymbol{\mu}_{\mathcal{P}|\mathcal{Q}}).
\end{equation}
When the mask is designed as $[\boldsymbol{\Delta}_{\mathcal{P}|\mathcal{Q}}]_{i^{\ast}i^{\ast}} = 1 $ and $ [\boldsymbol{\Delta}_{\mathcal{P}|\mathcal{Q}}]_{jj} = 0, \forall j \neq i^{\ast} $ for $i^{\ast} = \arg \max \boldsymbol{\mu}_{\mathcal{P}|\mathcal{Q}}$, it indicates that only the beam with the maximum mean of RSRP is considered. In this work, we propose a heuristic mask design as
\begin{equation}\label{equ:mask_app}
h({\bf a}) = \textup{sigmoid}\left(\beta ({\bf a} - \max {\bf a} + \alpha)\right),
\end{equation}
where $ {\bf a} $ is an input vector with element scalar $ a $, $ \textup{sigmoid}(a) = \frac{1}{1 + \exp(-a)} $, $ \alpha $ and $ \beta $ respectively are defined as the threshold and fairness coefficients. 

\subsection{Iterative Beam Prediction and Probing Beam Selection}\label{sec:iteration}

In this part, we give an iterative beam prediction and probing beam selection algorithm to solve the problems \eqref{equ:mle} and \eqref{equ:det_sigma}.

The primitive beam prediction problem \eqref{equ:mle} and the combinatorial optimization problem \eqref{equ:det_sigma} are intractable, especially when the beam space $ N $ is large. First, the estimation of the covariance matrix $ \boldsymbol{\Sigma} $ is difficult, requiring a large amount of offline data and having a high computational complexity $ \mathcal{O}(N^3) $. Besides, learning to generate a covariance matrix involves the matrix inversion operation in \eqref{equ:h_variable}, which easily causes numerical stability issues in engineering. Second, to obtain the optimal solution of the combinatorial optimization problem \eqref{equ:det_sigma}, the computational complexity of the exhaustive search is $ \mathcal{O}(C_N^{|\mathcal{Q}|} (N - |\mathcal{Q}|)^3) $, which causes a prohibitive overhead.

To address these issues and obtain a low-complexity feasible alternation, one way is to reduce the covariance matrix $ \boldsymbol{\Sigma} $ to be a diagonal matrix $ \boldsymbol{\Lambda} $, and the subsequent problem \eqref{equ:det_sigma} is simplified to a trivial question where the optimal combination is a subset of the first $ L $ indices with top diagonal values in $ \boldsymbol{\Lambda} $. However, this reduction completely ignores the correlations across beams, resulting in non-negligible performance degradation.

To achieve a better tradeoff between performance and computational complexity, we propose to alternatively and iteratively address the problems \eqref{equ:mle} and \eqref{equ:det_sigma} in a greedy manner. The number of iterations is equal to $ L $. The covariance matrix $ \boldsymbol{\Sigma} $ is simplified to a diagonal matrix $ \boldsymbol{\Lambda} $. During each iteration, we learn and estimate $ \boldsymbol{\Lambda}_{\mathcal{P} | \mathcal{Q}} $ using an iteration-specific variance network, instead of the closed-form solutions in \eqref{equ:close_1} and \eqref{equ:close_2}. Subsequently, we select only one probing beam by $ \boldsymbol{\Lambda}_{\mathcal{P} | \mathcal{Q}} $. Although the conditional variance $ \boldsymbol{\Lambda}_{\mathcal{P} | \mathcal{Q}} $ can be deducted with $ \boldsymbol{\Lambda} $, the estimation of $ \boldsymbol{\Lambda} $ is coarse and thus the estimation error will propagate with the deduction. The proposed retraining of the variance network can reduce the error propagation issue. Specifically, in the $ l $-th iteration, the beam prediction problem \eqref{equ:mle} is rewritten as
\begin{equation}\label{equ:mle_sim}
\max_{\boldsymbol{\Theta}_f^l, \boldsymbol{\Theta}_g^l} \mathbb{E}_{{\bf x}, {\bf s}, q^l} \Big\{ \ln \mathcal{N}({\bf x}_{\mathcal{P}^l}; \boldsymbol{\mu}_{\mathcal{P}^l | \mathcal{Q}^l}, \boldsymbol{\Lambda}_{\mathcal{P}^l | \mathcal{Q}^l}) \Big\},
\end{equation}
where
\begin{align}
\boldsymbol{\mu}_{\mathcal{P}^l | \mathcal{Q}^l} & = f({\bf x}_{\mathcal{P}^l | \mathcal{Q}^l}, \mathcal{Q}^l, {\bf s}; \boldsymbol{\Theta}_f^l),\label{equ:f_sim}\\
\boldsymbol{\Lambda}_{\mathcal{P}^l | \mathcal{Q}^l} & = g(\mathcal{Q}^l, {\bf s}; \boldsymbol{\Theta}_g^l),\label{equ:g_sim}
\end{align}
and $ \mathcal{P}^{l} = \mathcal{P}^{l-1, \ast} \backslash \{q^l\}, \mathcal{Q}^l = \mathcal{Q}^{l-1, \ast} \bigcup \{q^l\}, \forall q \in \mathcal{P}^{l-1, \ast} $ with $ q^l $ being the candidate probing beam index for selection in this round, $ \mathcal{P}^{l-1, \ast} $ and $ \mathcal{Q}^{l-1, \ast} $ respectively are the set of candidate probing beam indices and the set of probing beam indices in the previous round. In the initial round, $ \mathcal{P}^{0, \ast} = \{1, \cdots, N\}, \mathcal{Q}^{0, \ast} = \varnothing $. Then, the combinatorial optimization problem \eqref{equ:det_sigma} is simplified as a one-dimensional search problem as
\begin{equation}\label{equ:det_sigma_sim}
\min_{q^l \in \mathcal{P}^{l-1}} \det \boldsymbol{\Lambda}_{\mathcal{P}^l|\mathcal{Q}^l}.
\end{equation}
We denote the computational complexity of one inference of the variance network as $ \mathcal{O}(\omega) $. Then, the computational complexity of \eqref{equ:det_sigma_sim} is $ \mathcal{O}(\omega (N - l + 1)) $, and the total computational complexity of the greedy algorithm is $ \mathcal{O}(\omega \frac{L (2N - L + 1)}{2}) $. In the $ l $-th round, the selected probing beam index is represented as $ q^{l, \ast} $, $ \mathcal{Q}^{l, \ast} = \mathcal{Q}^{l-1, \ast} \bigcup \{q^{l, \ast}\} $, and $ \mathcal{P}^{l, \ast} = \{1, \cdots, N\} \backslash \mathcal{Q}^{l, \ast} $.

Considering the iterative solution with masking, the problem \eqref{equ:det_sigma_sim} is reformulated as
\begin{equation}\label{equ:det_sigma_mask_sim}
\min_{q^l \in \mathcal{P}^{l-1}} \det (\boldsymbol{\Delta}_{\mathcal{P}^l|\mathcal{Q}^l} \boldsymbol{\Lambda}_{\mathcal{P}^l|\mathcal{Q}^l}),
\end{equation}
where $ \boldsymbol{\Delta}_{\mathcal{P}^l|\mathcal{Q}^l} $ is the weight derived by \eqref{equ:mask_re} with $ \boldsymbol{\mu}_{\mathcal{P}^l|\mathcal{Q}^l} $.

\begin{algorithm}[h]
	\caption{Iterative beam prediction and probing beam selection (offline training).}
	\label{alg:alg_training}
	\KwIn{Dataset $ \mathcal{D} $, maximal number of probing beams $ L $.}
	\KwOut{Learned networks $ \{f^{l, \ast}, g^{l, \ast}\}_{l=1}^L $.}
	\BlankLine
	Initialize the set of candidate probing beam indices $ \mathcal{P}^{0, \ast} = \{1, \cdots, N\} $, the set of probing beam indices $ \mathcal{Q}^{0, \ast} = \varnothing $.
	
	\For{$ l = 1 $ to $ L $}{

	Choose $ \forall q^{l} \in \mathcal{P}^{l-1, \ast} $ randomly, and obtain $ \mathcal{P}^{l} = \mathcal{P}^{l-1, \ast} \backslash \{q^{l}\}, \mathcal{Q}^{l} = \mathcal{Q}^{l-1, \ast} \bigcup \{q^{l}\} $.
	
	Train the mean network $ f $ and the variance network $ g $ by \eqref{equ:mle_sim} with $ \mathcal{P}^{l}, \mathcal{Q}^{l} $.

	Select the probing beam index $ q^{l, \ast} $ by \eqref{equ:det_sigma_mask_sim} with learned $ f^{l, \ast} $ and $ g^{l, \ast} $.
	
	Update $ \mathcal{P}^{l, \ast} \leftarrow \mathcal{P}^{l-1, \ast} \backslash \{q^{l, \ast}\} $ and $ \mathcal{Q}^{l, \ast} \leftarrow \mathcal{Q}^{l-1, \ast} \bigcup \{q^{l, \ast}\} $.
	}
\end{algorithm}

\begin{algorithm}[h]
	\caption{Iter-BP\&PBS: beam selection for data transmission (online inference).}
	\label{alg:alg_inference}
	\KwIn{Relative MU location $ {\bf s} $, maximal number of probing beams $ L $.}
	\KwOut{Beam for data transmission $ \hat{i}^{\ast} $.}
	\BlankLine
	Initialize $ \mathcal{P}^{0, \ast} = \{1, \cdots, N\} $, $ \mathcal{Q}^{0, \ast} = \varnothing $, reload the learned mean and variance networks $ \{f^{l, \ast}, g^{l, \ast}\}_{l=1}^L $.
	
	\For{$ l = 1 $ to $ L $}{
		
		BS searches the optimal probing beam index $ q^{l, \ast} $ by \eqref{equ:det_sigma_mask_sim} with learned $ f^{l, \ast} $ and $ g^{l, \ast} $, and transmits the selected beam.
		
		MU reports the counterpart RSRP $ x_{q^{l, \ast}} $ to the BS.
		
		BS updates $ \mathcal{P}^{l, \ast} \leftarrow \mathcal{P}^{l-1, \ast} \backslash \{q^{l, \ast}\} $ and $ \mathcal{Q}^{l, \ast} \leftarrow \mathcal{Q}^{l-1, \ast} \bigcup \{q^{l, \ast}\} $.
	}
	BS estimates the RSRP with the mean network $ f^{L, \ast} $, and select the one with maximal predicted RSRP, i.e., $ \hat{i}^{\ast} $.
\end{algorithm}

\begin{figure*}[h]
	\centering
	\subfigure[Iter-BP\&PBS ($ L = 8 $). The probing beams are depicted in black squares.]{
		\includegraphics[width=7.0in]{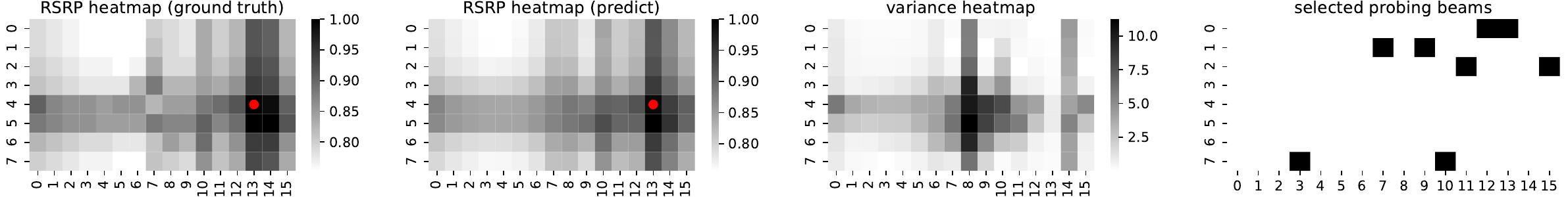}
		\label{fig:scheme_1_example}
	}
	\\
	\subfigure[2S-BP\&PBS ($ L_1 = 3, L_2 = 5 $). The probing beams at the first and second stages respectively are labeled in black and grey squares.]{		
		\includegraphics[width=7.0in]{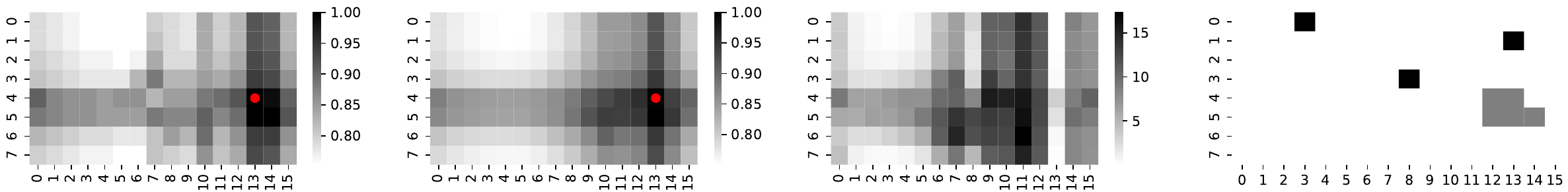}
		\label{fig:scheme_2_example}
	}
	\caption{Plots of selected probing beams and prediction results, where the x and y axes respectively are the numbers of antennas in the horizontal and vertical dimensions. The optimal beams are marked with red circles. The first sub-figure is the ground-truth of RSRP. The second and third sub-figures respectively are the mean and the corresponding variance estimates of the RSRP. The fourth sub-figure is the selected probing beams.}
	\label{fig:pbp_hat_example}
\end{figure*}

The offline collected training data is defined as $ \mathcal{D} = \{{\bf x}_j, {\bf s}_j\}_{j=1}^{N_{\textup{s}}} $ where $ N_{\textup{s}} $ is the number of samples. The parameter sets $ \boldsymbol{\Theta}_f $ and $ \boldsymbol{\Theta}_g $ are iteratively updated by mini-batch gradient descent (MBGD) until convergence. Fig.~\ref{fig:scheme_1_example} shows an illustrative ML prediction of RSRP. In summary, at the offline training stage, the iterative beam prediction and probing beam selection are given in Algorithm~\ref{alg:alg_training}, and the counterpart online inference is named as Iter-BP\&PBS and described in Algorithm~\ref{alg:alg_inference}. 

\begin{figure}[h]
	\centering
	\subfigure[Iter-BP\&PBS requires $ L $ information interactions. In the $ l $-th round, the BS selects and transmits the probing beam $ q^{l} $ by the measured beams $ {\bf x}_{\mathcal{Q}^{l-1}} $, then receives the counterpart RSRP report $ x_{q^{l}} $ and updates the measured beams as $ {\bf x}_{\mathcal{Q}^{l}} $.]{
		\centering
		\includegraphics[width=3.0in]{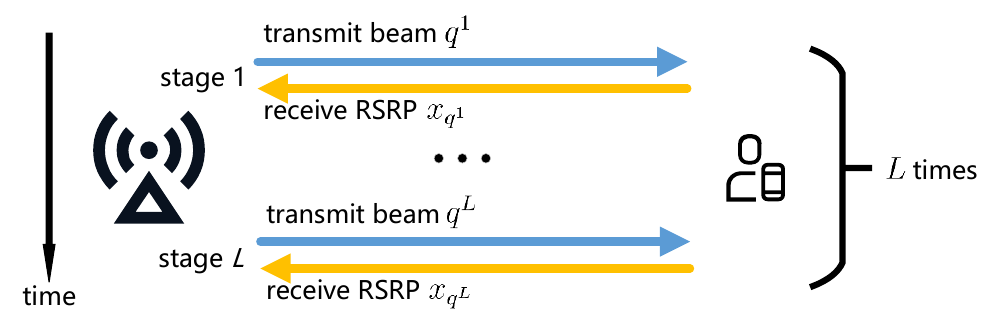}
		\label{fig:scheme_1}
	}
	\subfigure[2S-BP\&PBS requires $ 2 $ information interactions. At the first stage, the BS selects $ L_1 $ probing beams by the MU location, and receives the feedbacks $ {\bf x}_{\mathcal{Q}^{L_1}} $. At the second stage, the BS selects $ L_2 $ probing beams by $ {\bf x}_{\mathcal{Q}^{L_1}} $, and receives the feedbacks $ {\bf x}_{\mathcal{Q}^{L_2}} $.]{
		\centering	
		\includegraphics[width=3.0in]{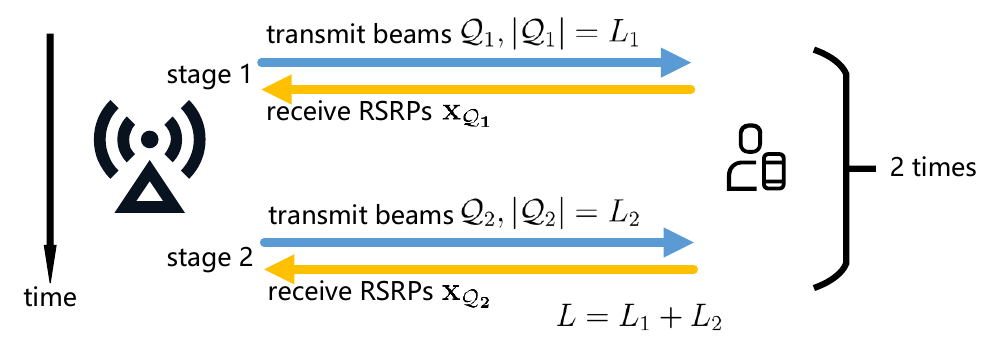}
		\label{fig:scheme_2}
	}
	\caption{Plots of information interactions between the BS and the MU.}
	\label{fig:schemes}
\end{figure}

\section{Two-stage Probing Beam Selection}\label{sec:two_stage}

In Section~\ref{sec:algorithm}, we have proposed an iterative probing beam selection and beam prediction algorithm, and we present a two-stage probing beam selection with much fewer information interactions in this section.

As shown in Fig.~\ref{fig:scheme_1}, the proposed Iter-BP\&PBS sequentially determines the training beams with RSRP feedback, is similar to the binary search with $ \log_2(N) $ interactions. The difference mainly is in the selection of the probing beams, where our proposed scheme decides by an entropy-based criterion, and the latter decides by binary comparison of RSRP values. Iter-BP\&PBS is still difficult to implement in practice for the following reasons.
\begin{itemize}
	\item \textbf{Interaction latency}. In each iteration, the mask in \eqref{equ:det_sigma_mask_sim} is updated by the RSRP feedback. Hence, the number of interactions between the BS and the MU is $ L $, and the latency linearly grows with the number of interactions.
	\item \textbf{Computational complexity}. In the $ l $-th iteration, the computational complexity is $ \mathcal{O}(\omega (N - l + 1)) $, so the total computational complexity of $ L $ iterations is $ \mathcal{O}(\omega \frac{L (2N - L + 1)}{2}) $.
\end{itemize}
Therefore, the significant interaction latency and computational complexity of Iter-BP\&PBS maybe be unacceptable for a real-time system. 

To address these issues, it is beneficial to design a location-aware probing codebook which is offline designed and online executed, and does not rely on the instantaneous feedback. As shown in Fig.~\ref{fig:scheme_2}, we propose a two-stage beam prediction and probing beam selection, i.e., 2S-BP\&PBS. Compared to Iter-BP\&PBS, the main revision is:
\begin{itemize}
	\item In Iter-BP\&PBS, the mask \eqref{equ:mask_re} is both location- and measurement-specific; but in 2S-BP\&PBS, the mask is only location-specific by approximating the input $ \boldsymbol{\mu}_{\mathcal{P}|\mathcal{Q}} \approx f({\bf s}) $.
\end{itemize}
This means that all probing beams can be fully determined by a location-aware codebook. However, without instantaneous measurement, the location-aware mask is not precise enough to guide the probing beam selection. Thus, we propose a two-stage probing method, where $ L_1 < L $ beams are measured at the first stage, the BS receives the feedbacks and subsequently decides on $ L_2 (L = L_1 + L_2) $ probing beams in the second stage. Moreover, the feedbacks of the $ L_1 $ beams experimentally are able to coarsely locate the strongest channel cluster. To further reduce the search complexity of the training beams in the second stage, we propose to select the top-$ L_2 $ beams w.r.t. the predicted RSRP for probing instead of the entropy-based beam selection.

Fig.~\ref{fig:scheme_2_example} shows an illustrative ML prediction of RSRP. In the following, we will respectively introduce the two-stage probing beam selection, i.e., codebook- and prediction-based beam probings.

\subsection{Codebook-based Beam Probing}

\begin{figure}[h]
	\centering
	\includegraphics[width=3.2in]{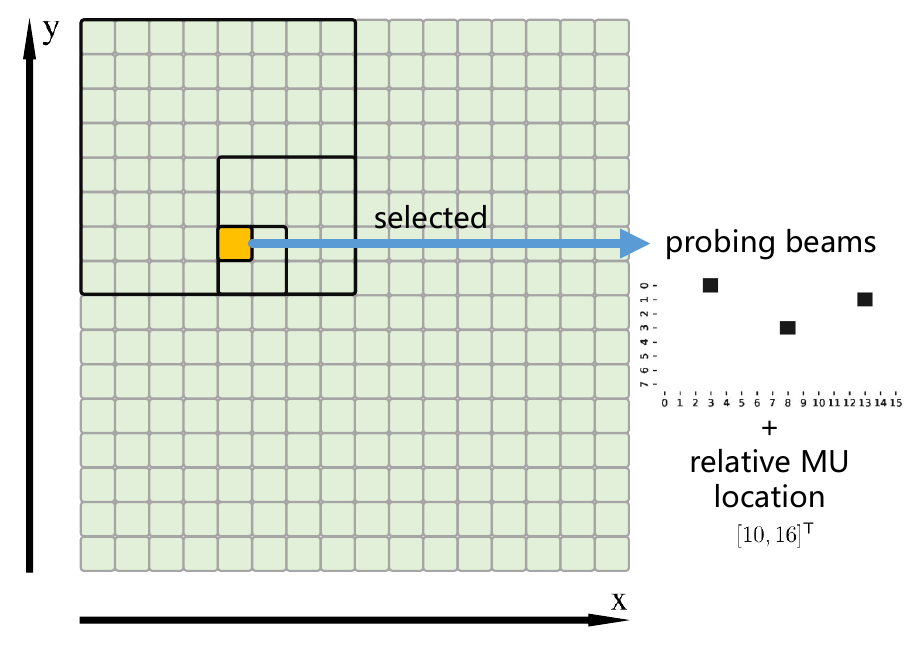}
	\caption{The 2D location-specific probing codebook $ \mathcal{C} $ for probing is composed of $ N_{\textup{x}} \times N_{\textup{y}} $ grids. Each grid stores a codeword including a probing beam set and a counterpart relative MU location. Given the location, the probing beams are selected by binary searching.}
	\label{fig:codebook}
\end{figure}

Codebook-based beam probing uses a location-specific codebook to generate probing beams. 

As shown in Fig.~\ref{fig:codebook}, the area covered by the BS is assumed to be a rectangle, and is evenly divided into $ N_{\textup{x}} \times N_{\textup{y}} $ square grids. The grid $ i, j $ stores a codeword including a probing beam set $ {\mathcal{Q}}_{i,j} $ and a counterpart location $ {\bf s}_{i,j} $, where $ i, j $ respectively are the indices of x-y coordinate. Regarding the $ l $-th element in $ \mathcal{Q}_{i, j} $, i.e., $ q^l_{i, j} $, is obtained by \eqref{equ:det_sigma_mask_sim}, where $ \boldsymbol{\mu}_{\mathcal{P}|\mathcal{Q}} $ to compute $ \boldsymbol{\Delta}_{\mathcal{P}^l|\mathcal{Q}^l} $ in \eqref{equ:mask_re} is approximated by $ f({\bf s}_{i, j}) $\footnote{Noting that $ \boldsymbol{\Lambda}_{\mathcal{P}^l|\mathcal{Q}^l} $ in \eqref{equ:det_sigma_mask_sim} remains unchanged.}. Thus, the probing codebook design is still entropy-based, and the corresponding procedure is summarized in Algorithm~\ref{alg:alg_training_2}.

During online inference, we propose to select the probing beams in $ \mathcal{C} $ with the closest corresponding distance $ d({\bf s}, {\bf s}_{i, j}), \forall i,j $ where $ d $ is a distance function. Formally, considering the Euclidean distance as $ d $, the selected probing beams are given by
\begin{equation}\label{equ:search}
	\begin{split}
		\mathcal{Q}^{L_1, \ast} & = \mathcal{Q}_{i, j} |_{i = i^{\ast}, j = j^{\ast}} \;  \textup{where}\\
		i^{\ast}, j^{\ast} & = \arg \min_j \|{\bf s} - {\bf s}_{i,j}\|_2^2, 1 \leq i \leq N_{\textup{x}}, 1 \leq j \leq N_{\textup{y}}.
	\end{split}
\end{equation}
The codebook $ \mathcal{C} $ is represented in the form of a binary tree, and the computational complexity of \eqref{equ:search} is $ \mathcal{O}(\log_2 N_\textup{x} N_\textup{y}) $ with binary search.

\begin{algorithm}[h]
	\caption{Probing codebook design.}
	\label{alg:alg_training_2}
	\KwIn{Dataset $ \mathcal{D} $, number of first probing beams $ L_1 $.}
	\KwOut{Probing codebook $ \mathcal{C} = \{\{\mathcal{Q}_{i,j}, {\bf s}_{i,j}\}_{i=1}^{N_{\textup{x}}}\}_{j=1}^{N_{\textup{y}}} $.}
	\BlankLine
	Initialize $ \mathcal{P}^{0, \ast} = \{1, \cdots, N\} $, $ \mathcal{Q}^{0, \ast} = \varnothing $, number of iterations $ L $, location-specific codebook $ \mathcal{C} = \{\{\mathcal{Q}_{i,j}, {\bf s}_{i,j}\}_{i=1}^{N_{\textup{x}}}\}_{j=1}^{N_{\textup{y}}} $ where $ \mathcal{Q}_{i,j} = \varnothing, \forall i,j $.
	
	Train the mean network $ f $ only with MU location.
	
	\For{$ l = 1 $ to $ L_1 $}{

		Choose $ \forall q^{l} \in \mathcal{P}^{l-1, \ast} $ randomly, and obtain $ \mathcal{P}^{l} = \mathcal{P}^{l-1, \ast} \backslash \{q^{l}\}, \mathcal{Q}^{l} = \mathcal{Q}^{l-1, \ast} \bigcup \{q^{l}\} $.
		
		Train the mean network $ f^l $ and the variance network $ g^l $ by \eqref{equ:mle_sim} with $ \mathcal{P}^{l}, \mathcal{Q}^{l} $.

		\KwOut{Learned networks $ f^{l, \ast} $ and $ g^{l, \ast} $.}
		
		Select the probing beam index $ q^{l, \ast} $ by \eqref{equ:det_sigma_mask_sim} with learned $ f^{l, \ast} $ and $ g^{l, \ast} $.
		
		Update $ \mathcal{P}^{l, \ast} \leftarrow \mathcal{P}^{l-1, \ast} \backslash \{q^{l, \ast}\} $ and $ \mathcal{Q}^{l, \ast} \leftarrow \mathcal{Q}^{l-1, \ast} \bigcup \{q^{l, \ast}\} $.
		
		\For{$ i = 1 $ to $ N_{\textup{x}} $}{
			
			\For{$ j = 1 $ to $ N_{\textup{y}} $}{
				
				Select the probing beam index $ q_{i, j}^{l, \ast} $ by \eqref{equ:det_sigma_mask_sim} with learned $ f^{l, \ast} $ and $ g^{l, \ast} $, where $ \boldsymbol{\mu}_{\mathcal{P}|\mathcal{Q}} $ in \eqref{equ:mask_re} is approximated by $ f({\bf s}_{i, j}) $.
				
				Update $ \mathcal{Q}_{i,j} \leftarrow \mathcal{Q}_{i,j} \bigcup \{q_{i, j}^{l, \ast}\} $.
			}
		}
	}
\end{algorithm}

\begin{algorithm}[h]
	\SetKwProg{Fna}{Stage 1}{}{}
	\SetKwProg{Fnb}{Stage 2}{}{}
	\caption{2S-BP\&PBS: two-level probing beam selection (online inference).}
	\label{alg:alg_inference_2}
	\KwIn{Relative MU location $ \hat{\bf s} $, numbers of first and second probing beams $ L_1, L_2 $.}
	\KwOut{Beam for data transmission $ \hat{i}^{\ast} $.}
	\BlankLine
	Reload the mean network $ f^{L_1, \ast} $, probing codebook $ \mathcal{C} $.
	
	\Fna{}{
		BS searches the nearest probing beams $ \mathcal{Q}^{L_1, \ast} $ in $ \mathcal{C} $ with $ {\bf s} $ by \eqref{equ:search}, and transmits the beams.
		
		MU reports the counterpart RSRPs $ {\bf x}_{\mathcal{Q}^{L_1, \ast}} $ to the BS.
	}
	
	BS estimates the RSRP $ \boldsymbol{\mu}_{\mathcal{P}^{L_1, \ast} | \mathcal{Q}^{L_1, \ast}} $ with the mean network $ f^{L_1, \ast} $.
	
	\Fnb{}{
		BS selects the top-$ L_2 $ beams by the predicted RSRP $ \boldsymbol{\mu}_{\mathcal{P}^{L_1, \ast} | \mathcal{Q}^{L_1, \ast}} $ as the second probing beams, i.e., $ \mathcal{Q}^{L_2, \ast} $
		
		MU reports the counterpart RSRPs $ {\bf x}_{\mathcal{Q}^{L_2, \ast}} $ to the BS.
	}
	
	BS replaces the corresponding prediction result with the measurement $ {\bf x}_{\mathcal{Q}^{L_2, \ast}} $, and selects the beam with maximal RSRP as the beam for data transmission, i.e., $ \hat{i}^{\ast} $.
\end{algorithm}

\subsection{Prediction-based Beam Probing}

Using the RSRP feedbacks of the codebook-based probing beams, prediction-based beam probing generates the second probing beams with top-$ L_2 $ predicted RSRPs.
  
In the first stage, the BS transmits the probing beams with indices $ \mathcal{Q}^{L_1, \ast} $ and receives the corresponding RSRPs $ {\bf x}_{\mathcal{Q}^{L_1, \ast}} $, predicts the RSRP $ \boldsymbol{\mu}_{\mathcal{P}^{L_1, \ast} | \mathcal{Q}^{L_1, \ast}} $ with the mean network $ f^{L_1, \ast} $. Using the RSRPs $ {\bf x}_{\mathcal{Q}^{L_1, \ast}} $, accurate estimation of the optimal beam poses a challenge for the predictor. However, the predictor is still capable to roughly locate the optimal beam. Hence, we directly select the top-$ L_2 $ beams from the predicted RSRP $ \boldsymbol{\mu}_{\mathcal{P}^{L_1, \ast} | \mathcal{Q}^{L_1, \ast}} $ as the second probing beams, i.e., $ \mathcal{Q}^{L_2, \ast} $. The MU then reports the counterpart RSRPs $ {\bf x}_{\mathcal{Q}^{L_2, \ast}} $ to the BS. The BS replaces the corresponding prediction result with the measurement $ {\bf x}_{\mathcal{Q}^{L_2, \ast}} $, and selects the beam with the maximum RSRP for data transmission, i.e., $ \hat{i}^{\ast} $.

In summary, the procedure of the 2S-BP\&PBS is clarified as follows. In the first interaction, the BS is aware of the MU location, selects $ L_1 $ beams from the codebook $ \mathcal{C} $ for probing, and receives the RSRP feedbacks from the MU. In the second interaction, the BS predicts the RSRP in beamspace with the MU location and the RSRP feedbacks, selects the top-$ L_2 $ beams as the probing beams to re-measure the RSRP, and decides the beam for data transmission with the measurement. The proposed 2S-BP\&PBS only has twice interactions and a computational complexity $ \mathcal{O}(\omega + \log_2 N_\textup{x} N_\textup{y}) \approx \mathcal{O}(\omega) $, and the corresponding online inference is given in Algorithm~\ref{alg:alg_inference_2}.

\section{Deep Learning-enabled Beam Predictor}\label{sec:transformer}

In Section~\ref{sec:iteration}, we have generally proposed the DL-enabled mean and variance networks, i.e., $ f $ and $ g $, and we will present the details in this section. 

\begin{figure*}[h]
	\centering
	\includegraphics[width=5in]{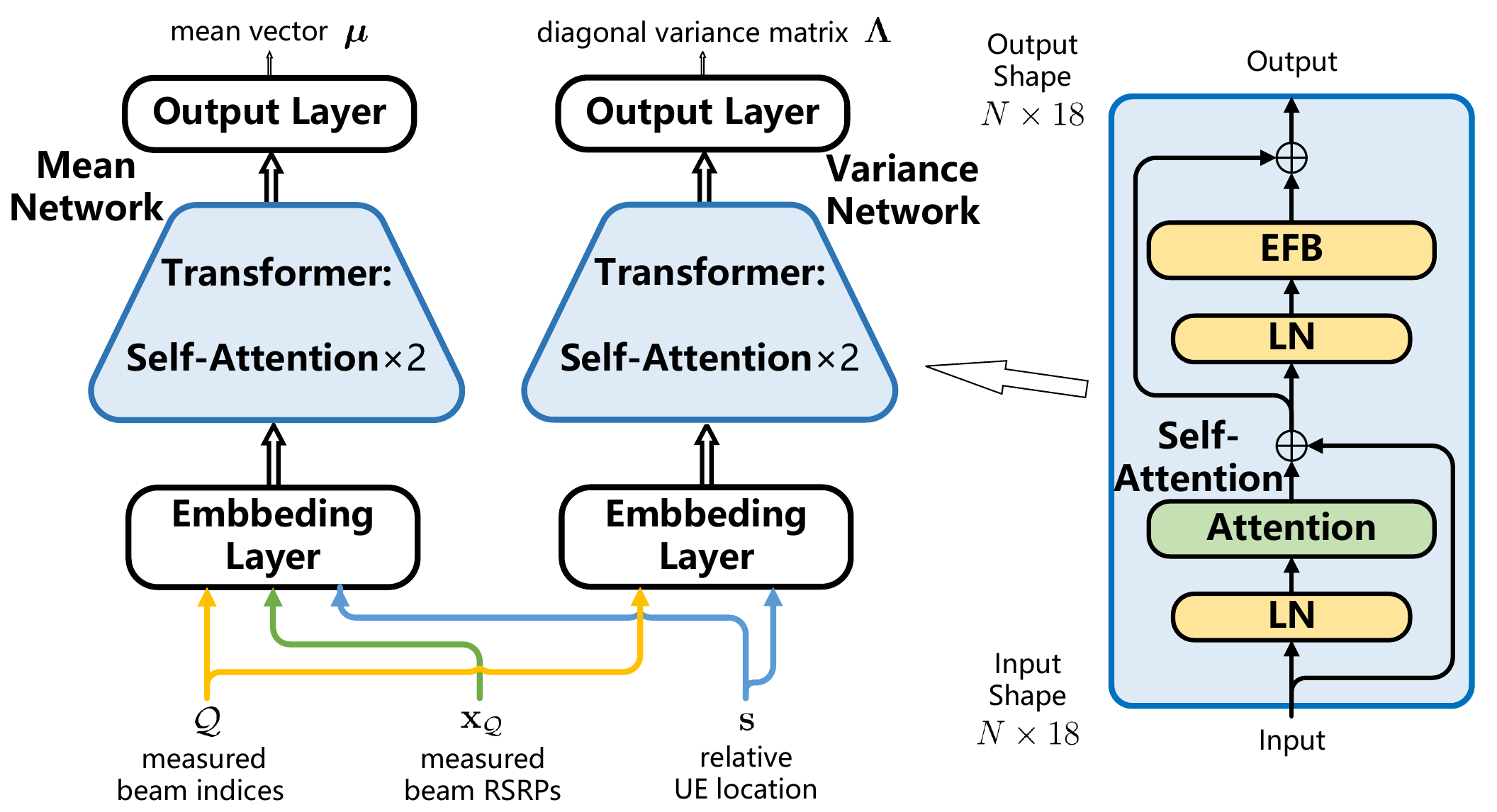}
	\caption{An illustration of the mean and variance networks (left) and a transformer (right).}
	\label{fig:transformer}
\end{figure*}

To achieve formidable learning capabilities, we design $ f $ and $ g $ with the transformer~\cite{attention} which is a transduction model that relies on self-attention to compute representations of its input and output. As shown in Fig.~\ref{fig:transformer}, both the networks are composed of three sequential blocks, i.e., an embedding layer, a transformer, and an output layer. In this work, we focus on the designs of the embedding and the output layers, while the transformer is quite mature in the literature, so we directly use it as the backbone.

\subsection{Embedding and Output Layers}

In Section~\ref{sec:iteration}, the dimension of the input $ {\bf x}_{\mathcal{Q}} \in \mathbb{R}^{|\mathcal{Q}| \times 1} $ is a variable w.r.t. the number of measured beams $ |\mathcal{Q}| $. To design a network scalable to $ |\mathcal{Q}| $, we propose to equivalently transform $ {\bf x}_{\mathcal{Q}} $ into $ {\bf x}'_{\mathcal{Q}} \in \mathbb{R}^{N \times 2} $ where the $ i $-th row vector $ {\bf x}'_i $ is
\begin{equation}
{\bf x}'_i = \begin{cases}
[x_q, 1],   & i \in \mathcal{Q},\\
[0, \;\, 0],     & \textup{otherwise},
\end{cases}
\end{equation}
where the first element is the RSRP value, and the second element indicates whether the corresponding beam is selected or not, to distinguish the measured and unmeasured beams both having zero RSRP values. Then, $ {\bf x}'_{\mathcal{Q}} $ is projected as $ {\bf x}''_{\mathcal{Q}} \in \mathbb{R}^{N \times 16} $, with a one-dimensional convolution operator\footnote{In this work, we only use a kernel size of $ 1 $, to make the input and output dimensions be equal, and also increase or decrease the data channel.} and the RELU function, i.e., $ \tilde{a} = \max(0, a) $. The relative MU location $ {\bf s} $ is linearly projected as $ {\bf s}' \in \mathbb{R}^{N \times 1} $. Additionally, a bias $ \textup{cls} \in \mathbb{R}^{N \times 1} $, i.e., the class token in \cite{attention}, is introduced. $ {\bf x}''_{\mathcal{Q}} $, $ {\bf s}' $ and $ \textup{cls} $ are concatenated as an input in $ \mathbb{R}^{N \times 18} $ for the cascaded transformer. The above embedding for the mean network $ f $ is also applicable for the variance network $ g $. The only difference is that the map $ g $ in \eqref{equ:g_sim} is conducted without measured RSRPs, so the first column of $ {\bf x}'_{\mathcal{Q}} $ are all zero.

The output of the transformer, i.e., the input for the output layer, is in the space $ \mathbb{R}^{N \times 18} $, and the counterpart output is in the space $ \mathbb{R}^{N \times 1} $, with a one-dimensional convolution and a RELU.

\subsection{Transformer}

As shown in Fig.~\ref{fig:transformer}, the proposed transformer $ g $ is a stack of different operations, layers and modules. The layer normalization (LN) operation is used to speed up training by normalizing the data into a standard normal distribution. In an expansion forward block (EFB), the input is linearly projected into an expanded space, after an activation layer, the expanded vector is projected back into the primary space. Residual connection is used to solve the training loss degradation problem in very deep networks by introducing an identity map. In Fig.~\ref{fig:transformer}, the residual connection operations are denoted by $ \bigoplus $. 

The core module in a transformer is self-attention, which is an attention mechanism that relates different positions of a single sequence to compute a representation of the sequence. An attention function can be described as mapping a query and a set of key-value pairs to an output, where the query, keys, values, and output are all vectors. The output is computed as a weighted sum of the values, where the weight assigned to each value is computed by a compatibility function of the query with the corresponding key. Given the input matrix $ {\bf A} \in \mathbb{R}^{M \times \tilde{M}} $, the key, query and value vectors are
\begin{subequations}\label{equ:qka}
	\begin{align}
	{\bf Q} & = {\bf W}_{\textup{q}} {\bf A},\\
	{\bf K} & = {\bf W}_{\textup{k}} {\bf A},\\
	{\bf V} & = {\bf W}_{\textup{v}} {\bf A},
	\end{align}
\end{subequations}
where superscript $ \tilde{(\cdot)} $ denotes the output marker, $ {\bf W}_{\textup{q}} $, $ {\bf W}_{\textup{k}} $, $ {\bf W}_{\textup{v}} $ respectively are the corresponding trainable linear transformation matrices, $ {\bf W}_{\textup{q}} $ and $ {\bf W}_{\textup{k}} $ both are in $ \mathbb{R}^{N_{\textup{d}} \times M} $ where $ N_{\textup{d}} $ is the feature dimension of the key matrix, and $ {\bf W}_{\textup{v}} \in \mathbb{R}^{M \times M} $. The result of the scaled dot-product attention is
\begin{equation}\label{equ:attention}
\tilde{{\bf A}} = {\bf V} \textup{softmax}\left(\frac{{\bf K}^{\mathsf{T}} {\bf Q}}{\sqrt{N_{\textup{d}}}}\right),
\end{equation}
where softmax:$ \tilde{{\bf a}} = \frac{\exp({\bf a})}{\sum_j \exp(a_j)} $. Since the variance of the inner product of $ {\bf K} $ and $ {\bf Q} $ increases with increasing embedding size, the result of the product is scaled by $ \sqrt{N_{\textup{d}}} $.

\section{Simulations}\label{sec:sim}

\subsection{Configuration}\label{subsec:config}

\begin{figure}[htb]
	\centering
	\includegraphics[width=3.0in]{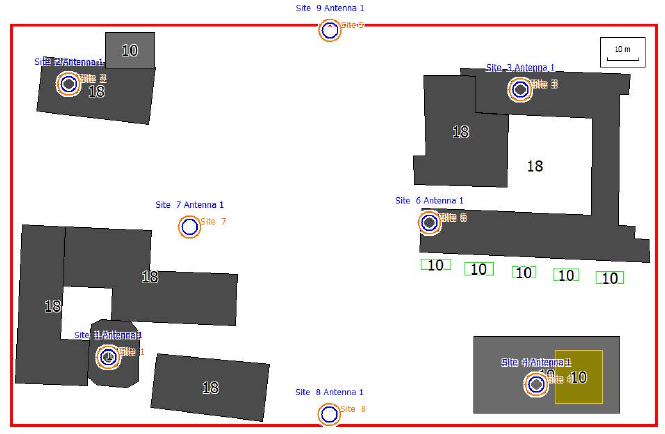}
	\caption{Layout of a mmWave communication scenario. The BSs marked in circles are located on the buildings or along the streets, and the MUs are distributed outdoors.}
	\label{fig:scene}
\end{figure}

To evaluate the performance of the proposed location-aware beam probing and prediction, which requires spatial consistency, the mmWave channel is established as a map-based hybrid model according to 3GPP 38.901 clause 8~\cite{38901}, consisting of deterministic and stochastic components. The deterministic ray tracing model is established by Feko Winprop~\cite{WinProp}. The consideration of random scattering by moving objects is used to verify the generalization ability of the proposed schemes. Each BS has 3 sectors covering the whole horizontal plane. The configurations of the BS and MU antennas are given in Table~\ref{tab:sim_config}, the feedback RSRP is logarithmically quantized with an accuracy of $ 1 \; \textup{dB} $. As shown in Fig.~\ref{fig:scene}, the geometric layout of the cell-free mmWave network covering $ 210 \times 130 \; \textup{m}^2 $ is a representative of the urban scenario. The BSs are located on top of the buildings or along the streets, and their heights are in the range $ [15, 40] $ m. The heights of buildings or trees are also marked. Data samples are uniformly collected from the outdoor area in Fig.~\ref{fig:scene}.

\begin{table}[htb]
	\centering 
	\setlength{\tabcolsep}{0.5mm}{
		\small
		\centering 
		\caption{Simulation Configurations of Scenario} 
		\begin{tabular}{c|c} 
			\toprule
			Name & Value\\ 
			\midrule
			carrier Frequency & $ 30 \; \textup{GHz} $ \\
			bandwidth $ B $ & $ 100 \; \textup{MHz} $\\
			number of BS antennas & $ 16 \times 8 $\\
			number of MU antennas & $ 1 $ \\
			symbol duration $ T_{\textup{s}} $ & $ 8.92 \; \mu s $\\
			time-slot duration $ T_{\textup{c}} $ & $ 20\; \textup{ms} $ \\
			noise power spectral density & $ -174 \; \textup{dBm/Hz} $\\ 
			maximal number of probing beams $ L $ & $ 8 $\\
			position noise variance $ \sigma_s^2 $ & $ 1 \; \textup{m}^2 $\\
			\bottomrule
		\end{tabular}
		\label{tab:sim_config}}
\end{table}

For performance evaluation, the following four key performance indicators are concerned.
\begin{itemize}
	\item \textbf{MSE}: The average prediction MSE (in $ \textup{dBm}^2 $) w.r.t. the RSRP in beamspace, i.e., $ \mathbb{E}_{{\bf x}, {\bf s}}\big\{\frac{1}{N} \sum_{i=1}^N (x_i - \hat{x}_i)^2 \big\} $. This is an indicator to evaluate the overall prediction error.
	\item \textbf{Top-$ K $ accuracy}: The ratio that the optimal beam is included in the top-$ K $ predicted beams (ranked by predicted RSRPs), where $ K \in \{1, 3, 5\} $.
	\item \textbf{RSRP difference}: The absolute RSRP difference (in $ \textup{dBm} $) between the predicted beam $ \hat{i}^{\ast} $ and the corresponding ground-truth $ i^{\ast} $, i.e., $ \mathbb{E}_{{\bf x}, {\bf s}}\big\{| x_{\hat{i}^{\ast}} - x_{i^{\ast}} | \big\} $.
	\item \textbf{Effective achievable rate (EAR)}: We define EAR as
	\begin{equation}
	\textup{EAR} \triangleq \mathbb{E}_{{\bf h}, {\bf n}_{x}}\left\{\left(1 - \frac{L T_{\textup{s}}}{T_{\textup{c}}}\right) \log_2 \left(1 + \frac{|{\bf A}[:, i^{\ast}]{\bf h}|^2}{\sigma_x^2}\right)\right\},
	\end{equation}
	where $ T_{\textup{s}} $ and $ T_{\textup{c}} $ respectively are the durations of a symbol and a time-slot. At the beginning of each time-slot, the probing beams are sent, each occupying one symbol resource.
\end{itemize}
The proposed beam predictor has already been illustrated in Fig.~\ref{fig:transformer}. The detailed hyper-parameters of training are listed in Table~\ref{tab:hyper_config}. The simulation platform is: Python 3.10, Torch 2.0.0, CPU Intel i7-9700K, and GPU Nvidia GTX 1070Ti. The following results are averaged over all BS in Fig.~\ref{fig:scene}, and the optimal results are highlighted in bold.
\begin{table}[htb]
	\centering 
	\setlength{\tabcolsep}{0.5mm}{
		\small
		\centering 
		\caption{Hyper-parameters} 
		\begin{tabular}{c|c} 
			\toprule
			Name & Value\\ 
			\midrule
			number of samples $ N_{\textup{s}} $ & 60,000\\
			number of epochs & $ 100 $ \\
			batch size & $ 200 $\\
			learning rate & $ 0.001 $\\
			number of transformer modules & $ 2 $ \\
			number of embedding channels & $ 16 $ \\
			number of multi-heads & $ 1 $ \\
			\bottomrule
		\end{tabular}
		\label{tab:hyper_config}}
\end{table}

\subsection{Cost Functions}\label{sec:cost_fun}

\begin{figure}[htb]
	\centering
	\includegraphics[width=3.5in]{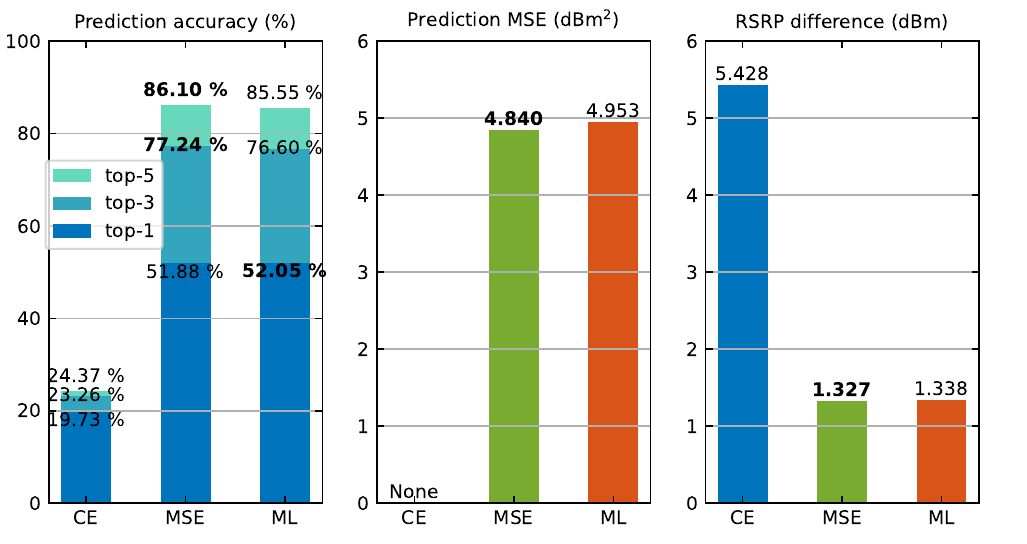}
	\caption{Predict performance of schemes with different cost functions.}
	\label{fig:uniform_compare}
\end{figure}

\begin{figure}[htb]
	\centering
	\includegraphics[width=2.5in]{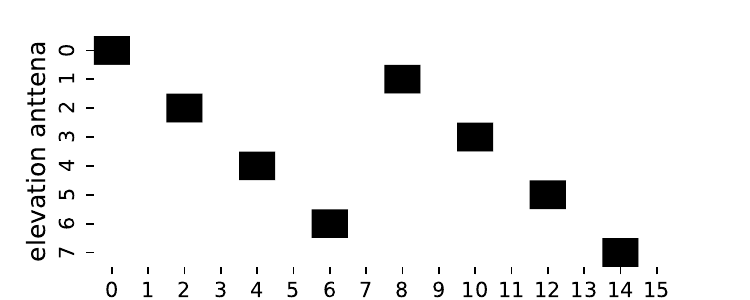}
	\caption{Baseline: uniform probing beams (plotted in black squares $ L = 8 $).}
	\label{fig:uniform}
\end{figure}

In this part, the prediction performance of schemes with different cost functions, i.e., cross entropy (CE) minimization\footnote{The activation function in the output layer is replaced by softmax.}, MSE minimization, and our proposed ML maximization, are studied. The prediction accuracy, MSE, and RSRP difference respectively are given in Fig.~\ref{fig:uniform_compare}. All schemes use the uniform probing beams plotted in Fig.~\ref{fig:uniform}. In general, the CE has a poor prediction result because it only learns the beam index with the maximum RSRP and neglects the RSRP in the whole beamspace. Meanwhile, the performance of MSE and ML is comparable. In the following studies, we use ML as the cost function to evaluate the prediction uncertainty.

\subsection{Learning Networks and Input Information}\label{sec:network}

\begin{figure}[htb]
	\centering
	\includegraphics[width=3.5in]{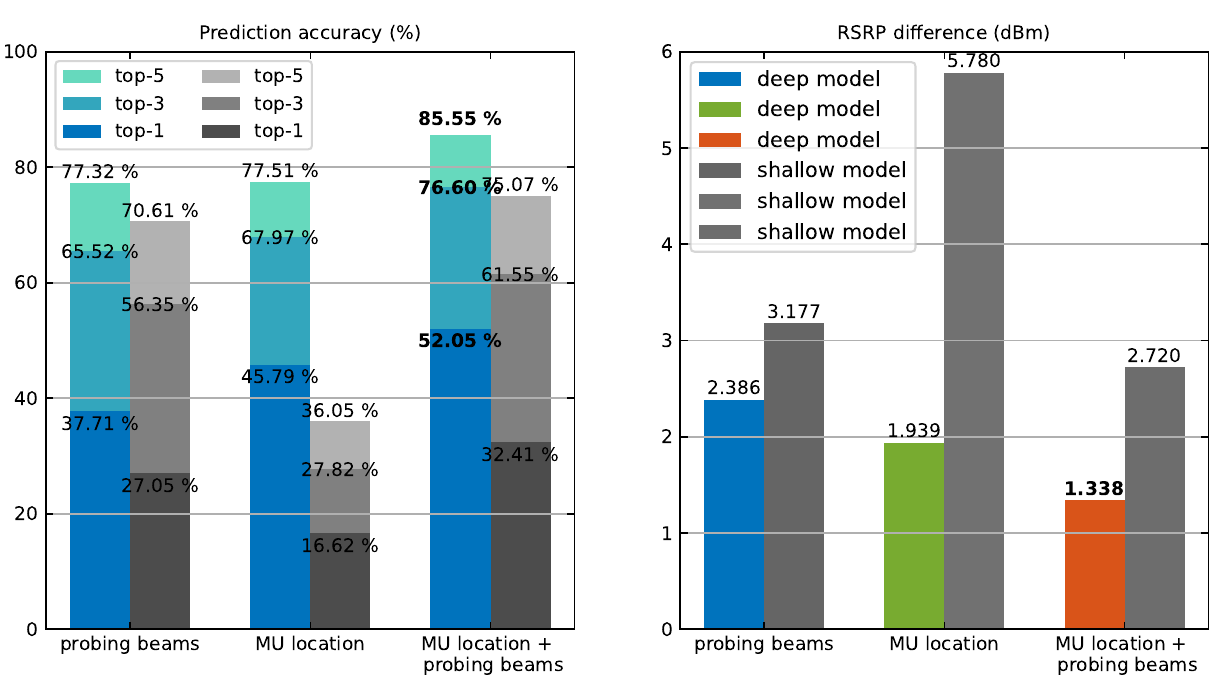}
	\caption{A comparison of prediction performance with different input combinations. The results with shallow and the proposed deep networks respectively are shown in grey and color.}
	\label{fig:information_compare}
\end{figure}

We then study the prediction performance with shallow and deep networks\footnote{GP is a typical shallow model, but it experimentally fails to work with the sklearn tool. Thus, we consider shallow NN as the shallow model.}, using different input combinations, i.e., 8 probing beams, MU location, 8 probing beams \& MU location. The deep network referred to the one proposed in Fig.~\ref{fig:transformer}, the shallow network is obtained by replacing the transformer backbone with 2 fully connected layers and each layer contains $ 128 $ neurons. As shown in Fig.~\ref{fig:information_compare}, the prediction performance of the proposed deep network (in color) significantly outperforms that of the shallow network (in grey), indicating that the potential map between the probing beams and/or MU location to the RSRP distribution is complex, and thus sufficient network depth can improve the prediction. 

\begin{figure*}[b]
	\centering
	\includegraphics[width=7.0in]{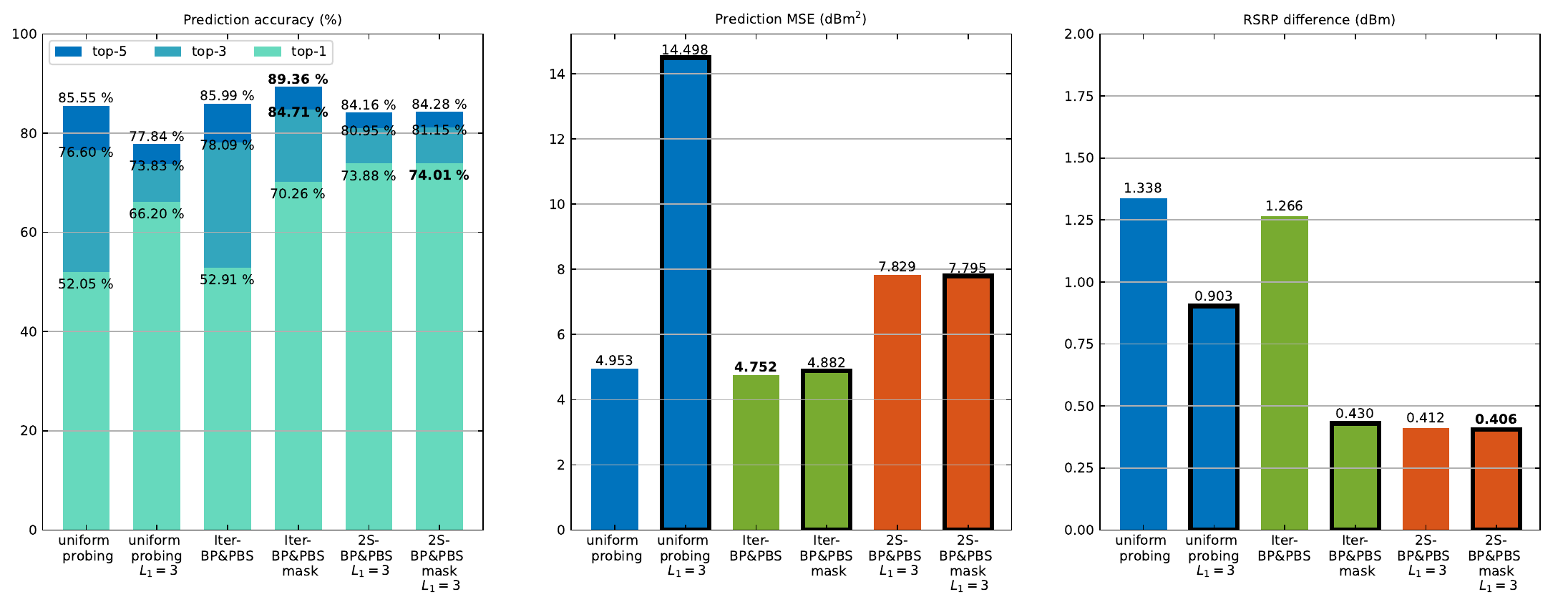}
	\caption{Prediction results of the Iter-BP\&PBS and 2S-BP\&PBS with/without mask.}
	\label{fig:uncertainty_compare}
\end{figure*}

On the other hand, considering different input combinations, the prior of MU spatial information significantly improves the RSRP prediction performance, and the top-$ 1 $ accuracy of MU location is up to $ 45.79\% $ which is better than $ 37.71\% $ of probing beams. This implies that location-aware probing beam-free BA/T is feasible if the requirement of RSRP difference is not critical. In the following, the scheme of the deep model with 8 probing beams \& MU location performs as a benchmark of uniform probing.

\subsection{Location-aware Beam Probing}\label{sec:site-specific}

The above subsections consider uniform probing beams, here we further discuss the influence of non-uniform probing beams. The location-specific probing beams can be designed using Iter-BP\&PBS or 2S-BP\&PBS, weighted or uniform entropy minimizations. In 2S-BP\&PBS, the coverage area is uniformly divided into $ 2 \times 2 \; \textup{m}^2 $ square grids.

As plotted in Fig.~\ref{fig:uncertainty_compare}, $ L_1 = 3 $ indicates that the 2S-BP\&PBS transmits probing beams twice, and the numbers of first and second probing beams respectively are $ L_1 = 3 $ and $ L_2 = 5 $. Compared to Iter-BP\&PBS, 2S-BP\&PBS significantly improves the top-$ 1 $ prediction accuracy and reduces the RSRP difference, by directly measuring the top-$ L_2 $ beams in the second interaction. Hence, the top-$ K $ ($ K > 1 $) accuracies of 2S-BP\&PBS are worse than those of Iter-BP\&PBS. Meanwhile, the proposed Iter-BP\&PBS and 2S-BP\&PBS significantly outperform the baseline with about 5 dB gain in RSRP difference.

In terms of prediction MSE with mask in \eqref{equ:mask_app}, the proposed Iter-BP\&PBS without mask has achieved the best result. This is because the use of weighted entropy minimization sacrifices the overall MSE performance to achieve a local entropy minimization on the areas with high predicted RSRP values. The mask design is consistent with the link-level BA/T whereas only the maximum RSRP is concerned. Moreover, our proposed schemes are also potential for the system-level BA/T optimizations, since we have predicted the RSRP in the whole beamspace and thus the estimation of the channel in the inference direction is feasible.

\begin{figure}[htb]
	\centering
	\subfigure[Maximal number of probing beams $ L $.]{
		\centering
		\includegraphics[width=3.2in]{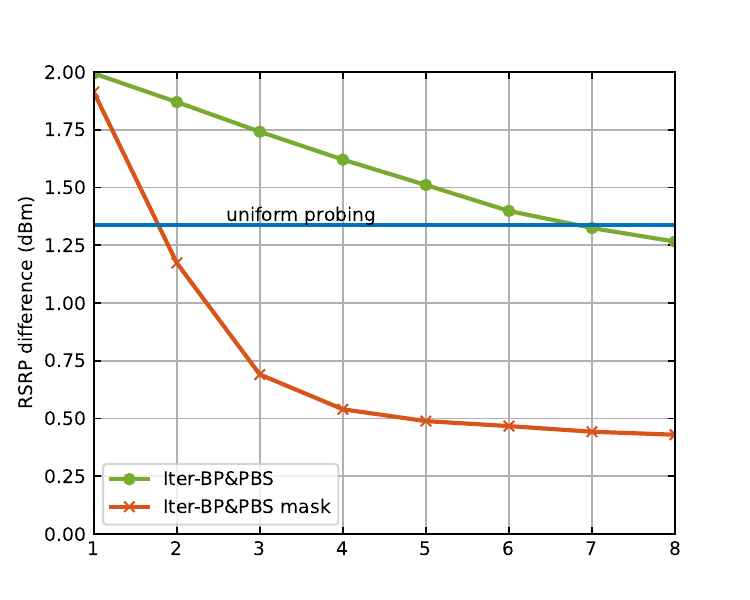}
		\label{fig:iteration_1}
	}%
	\\
	\subfigure[Number of first probing beams $ L_1 $ ($ L = 8 $).]{		
		\centering
		\includegraphics[width=3.2in]{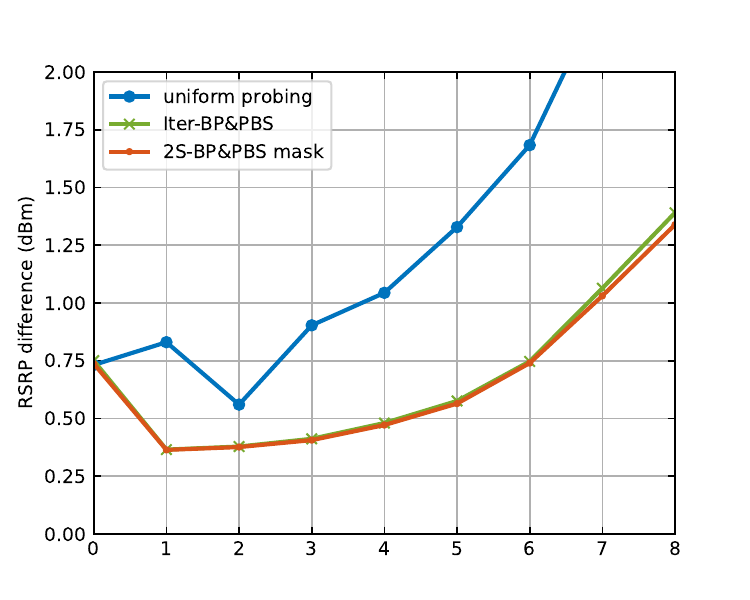}
		\label{fig:iteration_2}
	}%
	\centering
	\caption{RSRP difference versus number of probing beams.}
	\label{fig:iteration}
\end{figure}

Considering Iter-BP\&PBS, the RSRP difference versus maximal number of probing beams $ L $ is depicted in Fig.~\ref{fig:iteration_1}. When $ L \leq 3 $, the performance of Iter-BP\&PBS with the mask rapidly improves, then the improvement slows down as $ L $ continues to increase. Meanwhile, Iter-BP\&PBS without the mask has a bad performance, and it is comparable to the baseline when $ L = 7 $. This indicates that the proposed mask can help the beam predictor roughly locate the strongest cluster with a few probing beams, but an accurate prediction of the optimal beam is difficult. Thus, when the strongest cluster is roughly located, it would be better to re-design the mask in \eqref{equ:mask_app} or just probe the beams with the top-$ L_2 $ predicted RSRPs.

Considering 2S-BP\&PBS, the RSRP difference versus number of first probing beams $ L_1 $ is shown in Fig.~\ref{fig:iteration_2}. The uniform probing beams with $ L < 8 $ are designed so that the beams are uniformly located in beamspace. In particular, $ L_1 = 0 $ is the scheme that probes the top-$ L $ beams predicted by the networks with only MU location, and the beam probing at the first stage is removed, thus requiring an interaction. The scheme with $ L_1 = 8 $ is the one that probes the top-$ L $ beams predicted by the networks with only MU location, the beam probing at the second stage is removed and an interaction is also required. The scheme with $ L_1 = 0 $ is better than the one with $ L_1 = 8 $, indicating the importance of obtaining the optimal beam by measurement in the second stage. On the other hand, the proposed 2S-BP\&PBS significantly outperforms the baseline, and schemes with or without mask have similar performance, implying that the gain is mainly achieved by the entropy-based probing beam selection.

\subsection{Generalization Performance}\label{sec:generalization}

\begin{figure}[htb]
	\centering
	\includegraphics[width=3.2in]{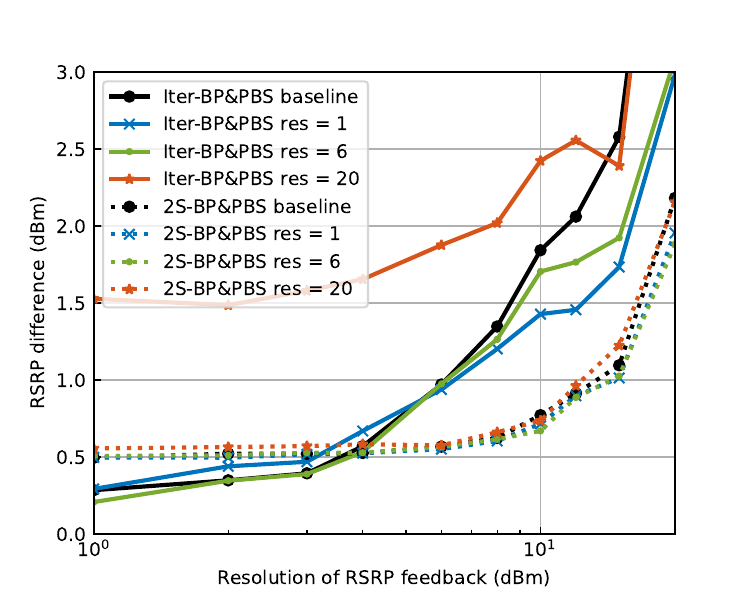}
	\caption{EAR versus SNR (number of users $ U = 1 $).}
	\label{fig:generalization_rsrp}
\end{figure}

In this subsection, we examine the generalization performance w.r.t. the quantization error of RSRP feedback.

The performance analysis of RSRP difference concerning the resolution of RSRP feedbacks is presented in Fig.~\ref{fig:generalization_rsrp}. RSRP values are constrained within the range of $ [-140, -40] $ dBm, and then quantized with resolutions ranging from $ 1 $ dBm to $ 20 $ dBm. The term `baseline' denotes the scheme of which both training and test data share the same RSRP resolution, while in other cases, the training data are generated with a fixed resolution. In the case of Iter-BP\&PBS, schemes with resolutions $ 1 $ and $ 6 $ dBm exhibit comparable performance to the baseline, showcasing good generalization. However, Iter-BP\&PBS struggles to predict accurately when the resolution becomes too coarse, specifically at $ 20 $ dBm. On the other hand, the proposed 2S-BP\&PBS demonstrates consistent performance across various resolutions, closely matching the baseline. This suggests superior generalization performance compared to Iter-BP\&PBS.

The impact of quantization error is akin to that of environmental white noise, and corresponding simulation studies are omitted due to constraints on article space.

\subsection{Data Transmission}\label{sec:data_transmission}

\begin{figure}[htb]
	\centering
	\includegraphics[width=3.2in]{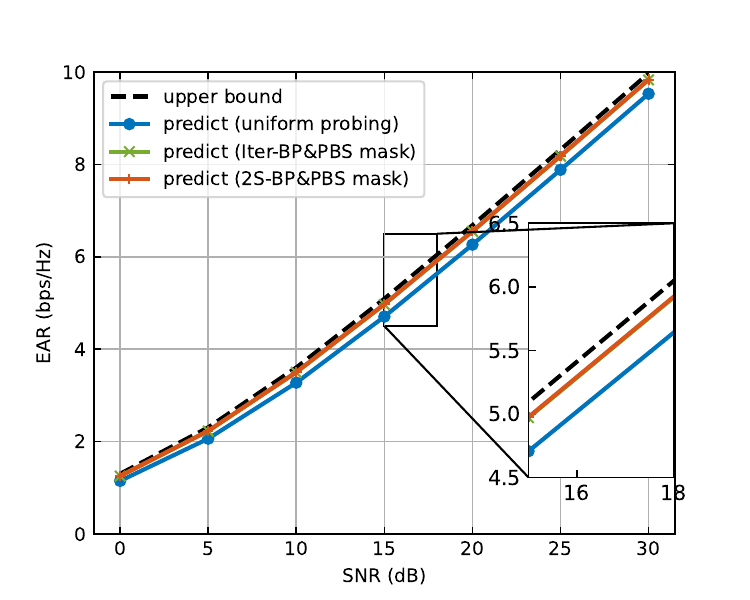}
	\caption{EAR versus SNR (number of users $ U = 1 $).}
	\label{fig:SNR}
\end{figure}

In this part, we study the EAR performance of data transmission with the proposed beam prediction schemes. The performance degradation of single-user BA/T is mainly determined by the beam alignment ratio, i.e., top-$ 1 $ accuracy. Regarding the influence of mis-alignment, there exists an EAR performance gap between the prediction-based schemes and the upper bound where the transmission beam is assumed to be correctly aligned without any training overhead. As shown in Fig.\ref{fig:uncertainty_compare}, the proposed schemes can align the transmission beam with a top-$ 1 $ accuracy about $ 70\% $, which is far from $ 100\% $ in a numerical sense. However, the mis-alignment does not mean that the transmission will fail to work, since the sub-optimal beams also have near-optimal RSRP. Thus, as shown in Fig.~\ref{fig:SNR}, the EAR performance versus signal-noise-ratio (SNR) of the proposed Iter-BP\&PBS and 2S-BP\&PBS is very close to the upper bound.

\begin{figure}[htb]
	\centering
	\includegraphics[width=3.2in]{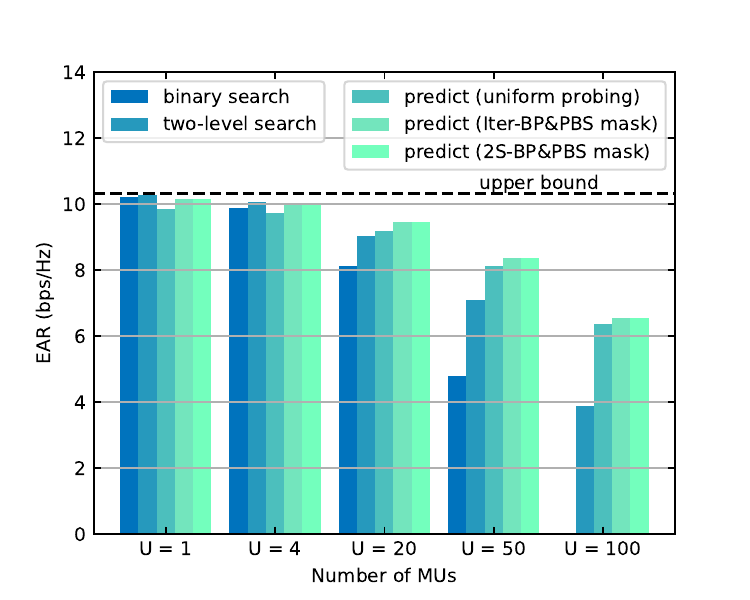}
	\caption{EAR versus number of users $ U $.}
	\label{fig:EAR}
\end{figure}

In terms of hierarchical measurement-based BA/T, the single-user overhead of two-level search is $ 24 $ including $ 16 $ wide beams and $ 8 $ narrow beams, and the overhead of binary search is $ 2 \log_2 N = 14 $ with $ 7 $ interactions. As we consider multiple MUs, the curves of EAR versus number of users $ U $ are demonstrated in Fig.~\ref{fig:EAR}. When $ U $ grows up to $ 100 $, all the time resources are consumed in the binary search, more than $ 60 \% $ time resources are consumed in the two-level search. Meanwhile, the prediction-based scheme has about $ 35 \% $ drop due to overhead cost.

\begin{table}[htb]
	\centering 
	\setlength{\tabcolsep}{0.5mm}{
		\small
		\centering
		\caption{Average computational time and storage costs} 
		\begin{tabular}{c|c|c|c|c} 
			\toprule  
			& \makecell[c]{two-level\\search} & \makecell[c]{binary\\search} & Iter-BP\&PBS & 2S-BP\&PBS\\ 
			\midrule
			storage cost (MB) & $ \backslash $ & $ \backslash $ & $ 25.152 $ & $ 3.144 $ + $ 0.321 $\\
			\makecell[c]{computational \\time cost (ms)} & $ \approx 0 $ & $ \approx 0 $ & $ 34.08 $ & $ 2.10 $\\
			\makecell[c]{number of \\interactions} & $ 2 $ & $ 7 $ & $ 8 $ & $ 2 $\\
			\bottomrule 
		\end{tabular} 
		\label{tab:complexity}}
\end{table}

Moreover, we list the average computational time and storage costs of the proposed schemes in Table~\ref{tab:complexity}. Considering the storage cost, $ 3.144 \; \textup{MB} $ and $ 0.321 \; \textup{MB} $ respectively are the storage space of a mean and variance network and a location-aware codebook. At the online inference stage, the mean network performs only one inference in 2S-BP\&PBS and the computational time cost is $ 2.10 \; \textup{ms} $, which roughly satisfies real-time deployment. Meanwhile, the computational time cost in Iter-BP\&PBS is $ 34.08 \; \textup{ms} $, which is greatly accelerated by the parallel computation in the GPU.

\section{Conclusions}\label{sec:conclusion}

In this work, we investigated the joint probing beam selection and probabilistic beam prediction, and formulated it as an entropy minimization problem. To solve this problem, we proposed an iterative scheme (Iter-BP\&PBS) with a simplified diagonal covariance matrix. To further reduce the number of interactions and the computational complexity of the Iter-BP\&PBS, we proposed a two-stage probing beam selection scheme, i.e., 2S-BP\&PBS. Simulation results demonstrated the superiority of the proposed schemes compared to the existing hierarchical beam search and beam prediction with uniform probing beams. In our future study, we will extend the entropy-based method and utilize the channel prior in frequency and time domains, for probing beam selection and beam prediction.

\bibliographystyle{IEEEtran}
\bibliography{References}

\end{document}